# Site-specific Deterministic Temperature and Humidity Forecasts with Explainable and Reliable Machine Learning

MengMeng Han[1], Tennessee Leeuwenburg[2*], Brad Murphy[2]

[1] Bureau of Meteorology, 32 Turbot St, Brisbane City, QLD 4000, Australia
[2] Bureau of Meteorology, 700 Collins St, Docklands, VIC 3008, Australia

## ABSTRACT

Site-specific weather forecasts are essential to accurate prediction of power demand and are consequently of great interest to energy operators. However, weather forecasts from current numerical weather prediction (NWP) models lack the fine-scale detail to capture all important characteristics of localised real-world sites. Instead they provide weather information representing a rectangular gridbox (usually kilometres in size). Even after post-processing and bias correction, area-averaged information is usually not optimal for specific sites. Prior work on site optimised forecasts has focused on linear methods, weighted consensus averaging, time-series methods, and others. Recent developments in machine learning (ML) have prompted increasing interest in applying ML as a novel approach towards this problem. In this study, we investigate the feasibility of optimising forecasts at sites by adopting the popular machine learning model gradient boosting decision tree, supported by the Python version of the XGBoost package. Regression trees have been trained with historical NWP and site observations as training data, aimed at predicting temperature and dew point at multiple site locations across Australia. We developed a working ML framework, named 'Multi-SiteBoost' and initial testing results show a significant improvement compared with gridded values from bias-corrected NWP models. The improvement from XGBoost is found to be comparable with non-ML methods reported in literature. With the insights provided by SHapley Additive exPlanations (SHAP), this study also tests various approaches to understand the ML predictions and increase the reliability of the forecasts generated by ML.

**KEYWORDS:** weather forecast; gradient boosting decision tree; machine learning; XGBoost; NWP post-processing; SHAP

## 1  Introduction

Despite continual improvements in Numerical Weather Prediction (NWP) over several decades, with high skill in forecasts of some parameters out to a week or more ahead, shortcomings in weather forecasts remain. These include model biases, random errors, and representativeness errors, and tend to grow with forecast lead time. Systematic biases can be corrected through post-processing (Vannitsem, et al. 2021), though random errors will remain so there will always be a level of uncertainty in a forecast. Deterministic, or single-value forecasts, produce one estimate of a forecast value, so ensembles of multiple forecasts, designed to capture the spread in possible outcomes, are increasingly being used to produce probabilistic forecasts (e.g. Richardson, 2000). These forecasts still typically represent a discrete area at least several kilometres in size, although this is decreasing as processing power grows. Model output at any current scale may not be representative of a specific site due to complex topography, land cover, urbanisation, or other local factors. A means of calibration between model forecasts and measured values at specific sites can improve this representation of specific locations. Methods for such calibration include those which alter forecast probability density functions to match that of the observed values (Bakker et al. 2019, Yang 2019, Alerskans and Kaas 2021).

---

\* Corresponding Author: tennessee.leeuwenburg@bom.gov.au, 700 Collins St, Bureau of Meteorology, Docklands VIC 3008, Australia

In addition to conventional statistical methods, various machine learning (ML) and deep learning (DL) models have also been experimented with for the task of site-specific weather forecasting. Site-specific weather forecasts are optimised for a specific individual weather observing station. The term 'local' weather forecast may refer to a site-specific forecast or to a small 'local' geographic area such as a suburb. Some of existing studies use time series of site observation data as the only input to generate forecasts for a single time step or multiple time steps. For example, XGBoost models have been applied to predict solar irradiance (Li et al. 2022); LSTM and its variations such as transductive LSTM (Karevan and Suykens 2020) and convolutional LSTM (Kong et al., 2022) have been used for temperature forecasts. These forecast models take site measurements of multiple relevant variables as input and formulate a multi-variate training dataset. Apart from purely observation data-driven models, ML can also be used as a post-processing approach for NWP - one which generates better local forecasts than NWP grid values. These models ingest both past NWP forecast and observation data as input, so the trained model can identify the systematic bias within the NWP model for the sites and generate a more accurate forecast once latest NWP data becomes available (Donadio et al. 2021; Hu et al. 2021; Sushanth et al. 2023).

Among all various ML models experimented in the literature, XGBoost (Chen et al. 2015) stands out for its computational efficiency, robustness and accuracy over alternative models. Previous studies have shown that XGBoost delivers better accuracy on tabular data, a data form often encountered in site-based forecasting. Grinsztajn et al (2022) argues that the better performance of tree-based models (XGBoost, Random Forest etc.) as compared to deep neural networks may be attributed to its robustness towards uninformative features and the fact that XGBoost does not suffer from the inductive bias toward smooth solutions as in the case of deep neural networks. Successful applications of XGBoost that yield satisfactory forecast accuracy have been reported for various scenarios and forecast variables, including temperature, precipitation (Done et al. 2023), wind power (Xiong et al. 2022), solar radiance (Li et al. 2022) and wave heights and periods (Zheng and Wu, 2019) etc. In these studies, the size of the training dataset varies from the scale of $10^3$ to $10^4$, and training time can be as low as less than 1.0s. The fast training of XGBoost enables hyperparameter tuning to optimize model accuracy, and allows for more frequent model updating and retraining, thereby utilising the latest available data in model training. Owing to the satisfactory performance of XGBoost as already benchmarked and experimented in previous studies, and its superiority of computational efficiency in training compared with deep learning models, XGBoost is selected for current study without further model selection.

With the constant development of bigger ML and DL models, and the growing size of the datasets used to train them, the complexity of the models has also been growing significantly. This has continued to the level that the internal structure of the model is no longer easily probed and the predictions of the models are more difficult to interpret. The low interpretability leads to complex models being considered as black-boxes and less trusted by customers, even if their accuracy has been validated on test datasets. The need to better understand, explain and interpret the outcome from ML and DL models results in growing recent research interest in the field of Explainable Artificial Intelligence (XAI) (Gunning et al. 2019; Linardatos et al. 2020). For weather forecasting models, it's also necessary to investigate and confirm that the causality learned by ML models agrees with that of the physics-based models and experience from operational weather forecasters. So far, various explanative methods have been proposed. On the scale of the entire dataset, it is possible to understand the ML model in terms of global feature importance. However, the disadvantage of feature importance is that it is calculated on the entire training dataset and cannot explain the reasoning of each individual output from the model. This prompts the proposal of local explanation methods, among which SHapley Additive exPlanations (SHAP) has been prominent. As a model-agnostic approach, SHAP has been used for local explanations for different model architectures and the insights provided by SHAP are also used for model debugging, model health monitoring and data drift detection (Lundberg et al. 2020; Duckworth et al. 2021). In the area of weather forecasting, the application of SHAP or other local explanation methods is still rare, with most of the studies only presenting the evaluation metrics of the trained models. However, recent studies have started to adopt SHAP for model explanations. For example, SHAP is used to explain spatial-temporal patterns picked up by the LSTM model for drought forecasts (Dikshit and Pradhan 2021a, 2021b) and the explanations are compared with physics-based models.

It should be noted that while SHAP helps explain the decision-making process towards an individual output, it does not provide insight into the correctness or accuracy of that output. The need to understand the quality of each prediction point from a ML model for better risk awareness and management prompts the studies of reliable machine learning. It is understood that the imperfections of a ML model can originate from the quantity and quality of data, and model architecture (Zhang et al. 2022). With insights into data properties and ML mechanisms, it is sometimes possible to identify those individual predictions with a higher risk to be misclassified or inaccurate and therefore less reliable than a global metric implies. The principles of pointwise reliability have been proposed by (Saria and Subbaswamy 2019), and various methodologies to evaluate reliability have been investigated by (Schulam and Saria 2019, d'Eon et al. 2022, Eyubogly et al. 2022). The real-time response mechanism to unreliable predictions is also integrated into some ML systems. For example, an abstention option is enabled in some classification models so that when a classification is identified to be unreliable, the model abstains without providing an output (Hellman 1970).

This paper presents the application of XGBoost as a post-processing approach for NWP grid data to produce better site-specific temperature and dew point forecasts. Forecast accuracy is evaluated both on common metrics and customized ones based on customer interests, and the results show that significant improvement can be achieved compared with original gridded data. The predictions generated by XGBoost are explained by SHAP so that additional insights into the trained model can be obtained. As a further step for reliable ML, the study also explores various ways to evaluate pointwise reliability of the forecasts, so that those parts of the predictions that are subject to higher risk of inaccuracy are identified at the time the prediction is made. To the authors' best knowledge, this is the first attempt to include the evaluation of pointwise reliability into the ML model for an NWP post-processing task.

## 2 Data overview

### 2.1 Gridded numerical data (IMPROVER)

The gridded numerical data used in this study is from the state-of-the-art probabilistic ensemble post-processing and verification system known as IMPROVER (Integrated Model post-PROcessing and VERification). IMPROVER ingests and blends raw forecasts from various NWP models, applies gridded calibration, and generates fully probabilistic forecasts (Roberts et al. 2023). The Australian version of IMPROVER developed by the Bureau of Meteorology has a grid covering the Australia region and adopts a cartesian coordinate system. The gridded data has a spatial resolution of 4.8km at screen level and 9.6km on upper levels. It generates hourly forecasts every 6 hours (4 issues a day) with lead hours up to 192. Figure 1(a) shows a snapshot of IMPROVER gridded forecast temperature at screen level on Aug 24, 2022 with lead hour 2.0, which shows the extent of the spatial coverage of IMPROVER data.

Since this current study focuses on deterministic forecasts only, the expected value of the IMPROVER probabilistic forecast at each location is extracted for training ML models[1]. In this study, the grid value of selected variables at 11 Australian sites (Figure 1(b)) up to lead day 7 are extracted from each issue forecast from IMPROVER, forming hourly time series as shown in Figure 1(c). The variables selected as features are: temperature at screen level ($T$), dew point temperature at screen level ($T_d$) and wind vector $u$ and $v$ at 10m height. Upper layer variables are not included since the spatial resolution of those variables is lower. Available IMPROVER data at time of this study ranges from Aug, 2022 to May, 2023. The sites selected for this study are scattered in different parts of Australia to ensure that ML model accuracy is not location dependent. Pointwise IMPROVER root-mean-square error (RMSE) of same-day forecasts on the selected sites in the studied period range from 1.2 ºC -2.3ºC for temperature, and 1.2 ºC - 2.9ºC for dew point temperature.

One important properties of IMPROVER forecast that affects ML training strategy is that the IMPROVER forecast error generally grows by lead hour. This is expected as NWP models derive much of their predictability from the initial conditions, and so gradually lose accuracy with longer forecast lead time. To quantify and visualize this effect, Figure 2(a) shows the IMPROVER temperature forecast of a single site $T_N$ in October with respect to the actual observed temperature for each forecast ($T_O$). The scatter plot shows that while numerical forecasts on both lead day 0 (2-24 hrs) and lead day 8 (168-192 hrs) exhibits a positive correlation with observed values, the forecast at lead day 0 is more accurate and less scattered. Figure 2(b) shows the distribution of forecast error for the same data points, which indicates that the forecast errors on lead day 8 have slightly higher absolute mean and twice as high standard deviation as compared with forecasts on lead day 0.

---

[1] Early investigations showed that taking various straightforward ensemble approach did not improve the skill of the forecast or properly capture the distribution. Further research is left for future work.

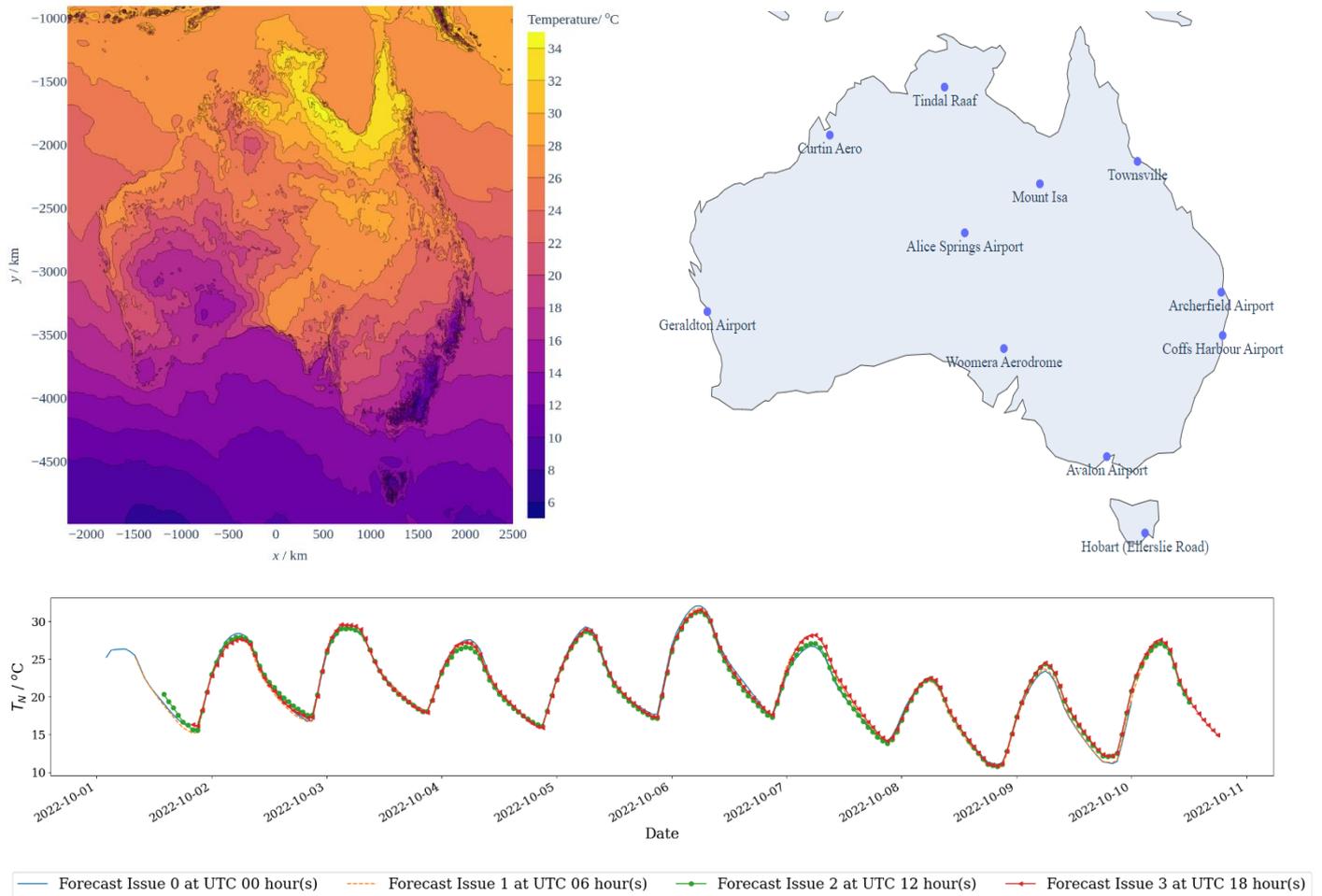

**Figure 1:** *(a) IMPROVER gridded data, (b) Selected sites and (c) Extracted daily time series from gridded data for Alice Springs. Each single-day forecast contains 4 issues (updates) published at UTC 0, 6, 12 and 18.*

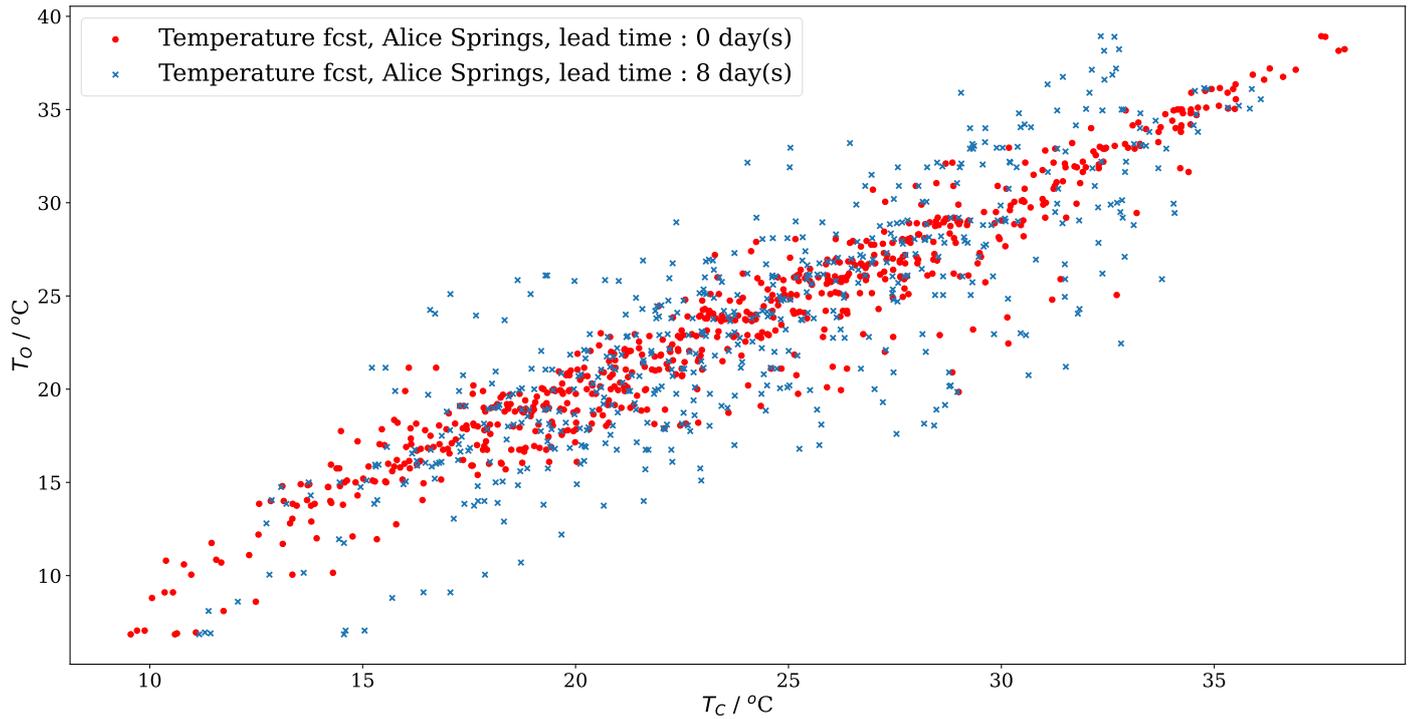

***Figure 2a:*** *Overview of IMPROVER hourly temperature forecast. Numerical value from IMPROVER $T_C$ vs. Observed site value $T_o$, temperature.*

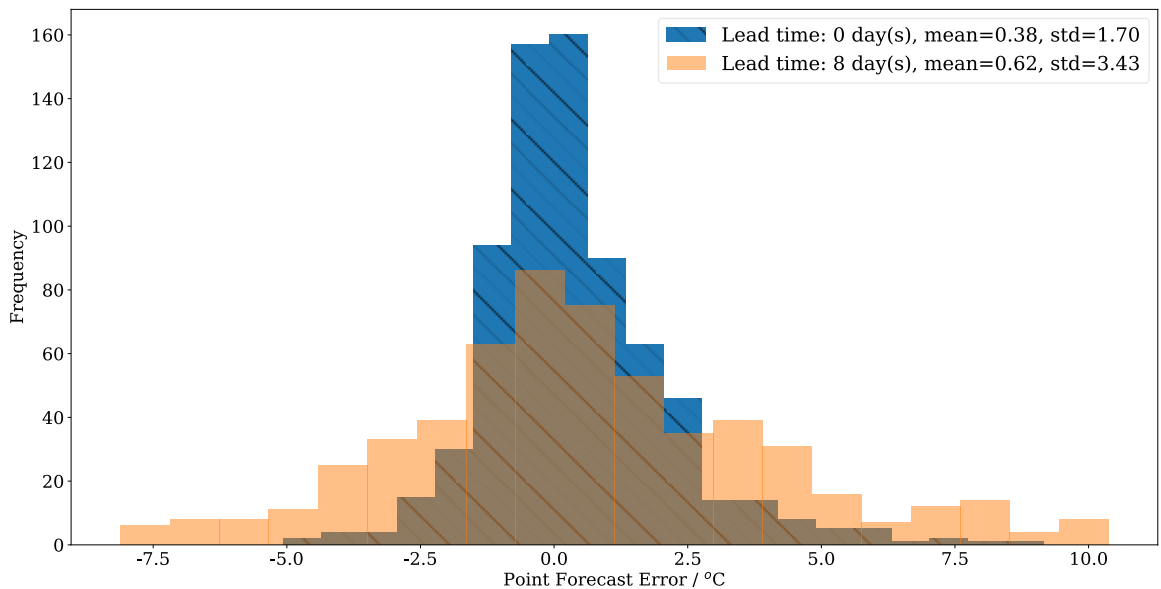

***Figure 2b:*** *Overview of IMPROVER hourly temperature forecast. Distribution of IMPROVER hourly temperature error, 1st-10th October 2022.*

## 2.2   Site observations

Observational data of selected sites are extracted from Australian Data Archive for Meteorology (ADAM) database. ADAM stores meteorological observations from Bureau of Meteorology managed observing systems over mainland Australia and from neighbouring islands, the Antarctic, ships and ocean buoys. The stored observation data is quality-controlled. While on-site measurements inevitably contain random error, the extracted temperature and dew point data is considered ground truth in this study. The underlying instruments collect data each minute, and the data are reprocessed into ten-minute mean values, valid on-the-hour.

## 3 Model training and optimisation

As stated in Introduction, XGBoost is selected for this study. The extracted time series data at each site location shown in Figure 1 can be directly used for training. However, to optimize the accuracy of the XGBoost model, proper pre-processing of the grid value from IMPROVER is necessary. Prior to training, the following pre-processing approaches are applied to the site data. It should be noted that the pre-processing techniques adopted in this study are not exhaustive and there are alternatives in the literature. The ones in this study are selected because they show the best performance in the parametric study.

- *Inclusion of surrounding grid values*: during data extraction from the IMPROVER gridded dataset, the grid value at the site location plus those around the location are extracted and used as separate input features. More specifically, for each of the site, the values on a grid are extracted. The grid is centred on the site location and each nearby point is 5 grid cells away as shown in Figure 3. Based on the orientation angle, the values of the 8 nearby points are marked by subscripts: NW, W, SW… etc, while the central value has the subscription of C. Feature names hereinafter follow this convention. As shown in the parametric studies below, the inclusion of these neighbouring cell values can greatly increase the model accuracy. These values are also physically important since the spatial variation of a variable indicates the instant spatial flux of that variable. The spacing between the neighbouring points are selected empirically, but a general rule applies that the points being too close leads to highly correlated features being added to the training set. On the other hand, the point being too remote leads to irrelevant features in the training set.
- *Data selection*: As shown in Figure 2(a), IMPROVER data with longer lead time can be considered as containing the same information but mixed with higher random error. Since good data quality in the training set is essential in building an accurate ML model, and data on lead day 0 already has four issues (shown in Figure 1(c) for augmentation, only data with lead time 0 is used for training. The model is expected to learn the essential data pattern that relates to the locational characteristics of the site, rather than the evolution of the forecast over time, and thus predictions of all lead days are generated using the same model.
- *Model inputs and outputs:* The input variables are: Air temperature at ground level (including surrounding grid points), wind at 10m above ground level at the station location, expressed as $U_{10}$ and $V_{10}$ components, and dewpoint temperature (including at surrounding grid points). The output variables are air temperature and dew point. An individual model is trained for each station and output (i.e. each model predicts a single variable for a single station).
- *Automatic feature selection*: In section 2.1, selected variables are extracted from IMPROVER grids as features. However, not all features are equally relevant for each site and may also vary between training runs. To remove irrelevant features in the training set, a pre-train with all features included is first performed for each station, and the 10 features with highest statistically global importance will be used for subsequent training while other less important features are removed. This proves to slightly increase the accuracy of the model and more importantly reduce the volume of data needed for validation and generating predictions.
- *Scaling*: XGBoost models do not perform extrapolation (Malistov and Trushin 2019) and the therefore may not generate accurate predictions with feature values that are outside the bounds of the training set. This is detrimental for the prediction of extreme values, such as new high records of temperature. To improve the performance, the original data is column-wise scaled to achieve a standard distribution for training and validation sets. Note for validation sets, the mean and variation is calculated based on the data of the last week of the training period instead of the whole training period. This matches the operational use case, whereby the forecast period is in the future and, as an unknown, the exact scaling factor cannot be known in advance. Using the most recent week aims to ensure that the calculated mean is the closest estimation of the upcoming validation period. This process may prevent inaccurate mean value applied on the validation set when daily average temperature is constantly increasing. With this process the bound of the two sets can mostly overlap with each other. The effect of scaling can be quantified in detail in subsequent Table 1.

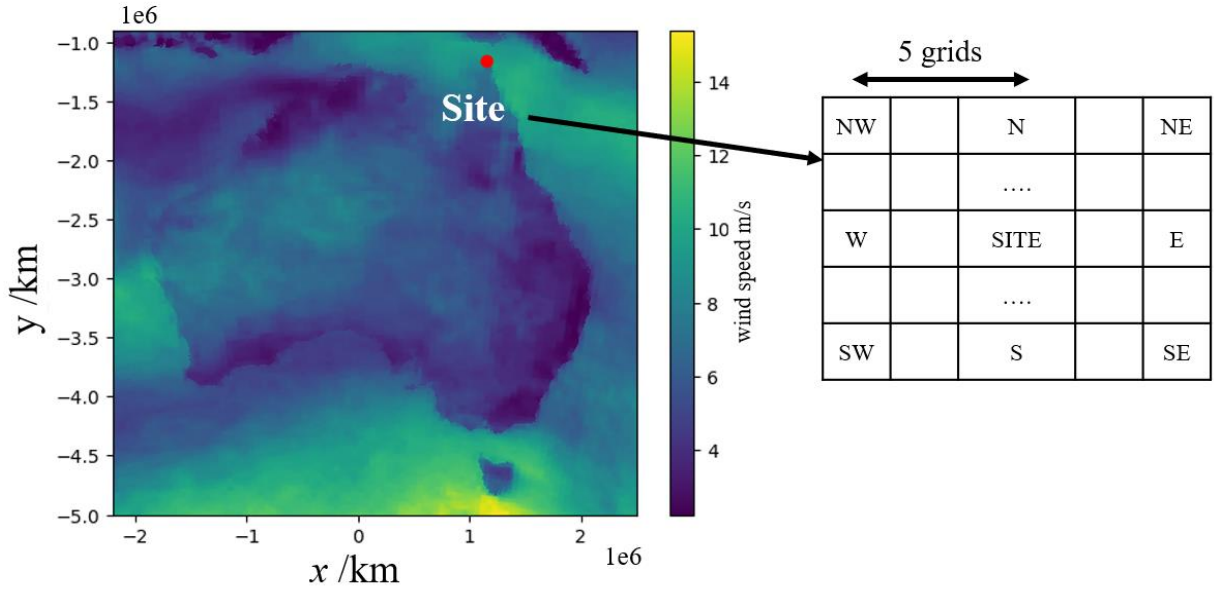

*Figure 3:* Positions of extracted grid values at and around a single site.

In addition, current study adopts a sliding window approach to continuously train new models on a 9 week training set and test on the subsequent one week. The advantage is that most recent weather patterns can be captured. Sliding window also utilizes the fast training enable by XGBoost to avoid data drifting due to infrequent model updates. Hyperparameters are tuned with standard 5-fold cross validation.

Table 1 shows a parametric study based on the temperature and predictions of one month period. Each column shows the resultant change in the average hourly RMSE caused by different data pre-processing settings. The change of RMSE is calculated by considering all lead times with equal weight, and the sum in each column is the aggregation of all changes from all tested sites. The results show that the prediction with all data pre-processing techniques applied has the lowest RMSE, and any of the preprocessing missing would lead to higher error.

*Table 1:* Change of RMSE ($^{o}$C) of temperature predictions averaged on all lead days for each site when different preprocessing approach is absent from data pipeline. Negative means better prediction accuracy. Prediction period: Oct, 27, 2022 – Nov, 27-2022.

| Sites/Preprocessing Methods | Change of RMSE, if feature selection is not performed | Change of RMSE, if surrounding points are excluded | Change of RMSE, if scaling is not performed |
|---|---|---|---|
| Alice Springs Airport | -0.01 | -0.09 | 0.50 |
| Archerfield Airport | 0.01 | 0.02 | 0.20 |
| Avalon Airport | 0.01 | -0.03 | 0.13 |
| Coffs Harbour Airport | 0.03 | 0.12 | 0.19 |
| Curtin Aero | 0.00 | 0.12 | 0.24 |
| Geraldton Airport | 0.01 | 0.19 | 0.09 |
| Hobart (Ellerslie Road) | -0.01 | 0.03 | 0.25 |
| Mount Isa | 0.01 | 0.04 | 0.68 |
| Tindal RAAF | -0.01 | -0.10 | 0.00 |
| Townsville | 0.02 | 0.09 | 0.28 |

| Woomera Aerodrome | -0.02 | -0.01 | 0.30 |
|---|---|---|---|
| SUM | 0.05 | 0.37 | 2.87 |
| AVERAGE | 0.005 | 0.034 | 0.261 |

As a first step towards a working prototype of a production system, a framework for automatically downloading, data pre-processing, ML training and validation framework is developed and named Multi-SiteBoost (MSB). The framework comprises three parts: (1) gridded IMPROVER data and site observation data are downloaded and stored. Data for the selected sites and features are then extracted from the large volume of data into time series; (2) An XGBoost model is automatically trained for each site and each sliding window in parallel; (3) The saved models are used to generate predictions, metric statistics and evaluating pointwise reliability on each of the predictions. Figure 4 shows the flow chart of MSB workflow. ML results as discussed in following sections are all generated by MSB framework.

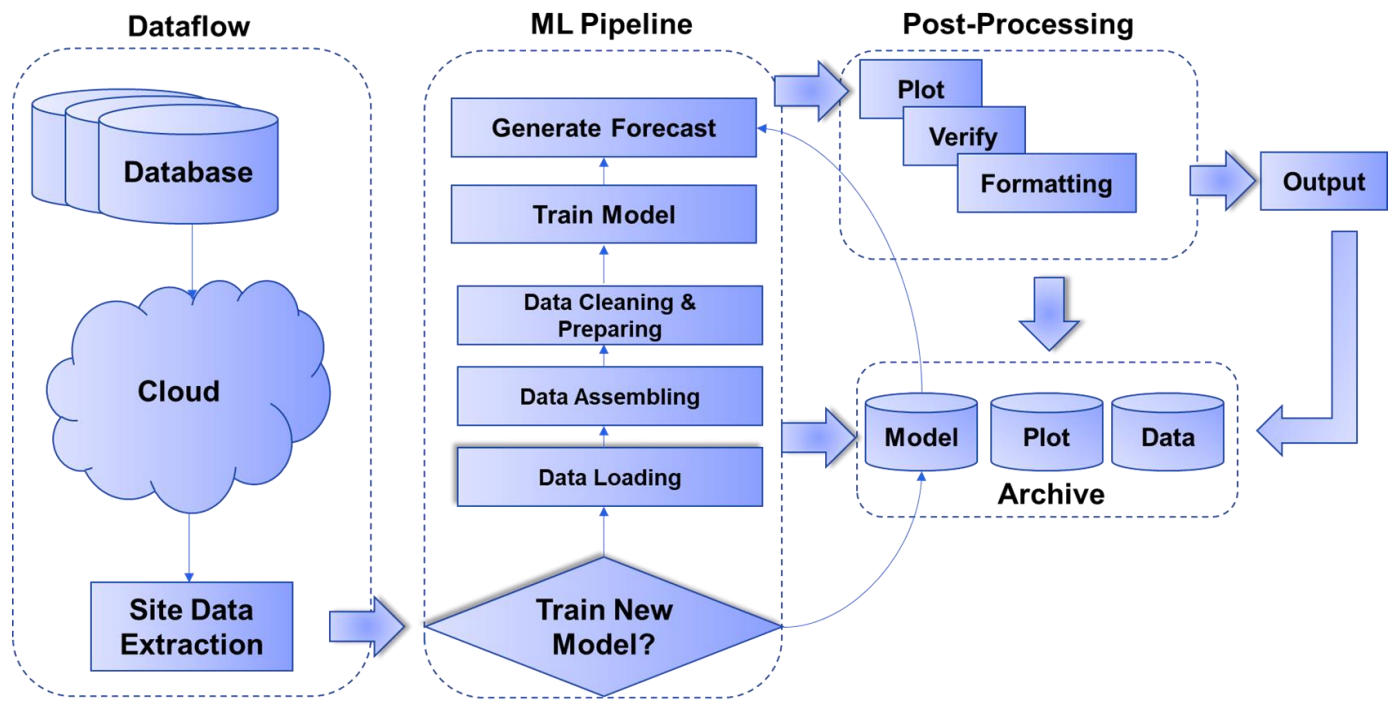

*Figure 4:* *Multi-SiteBoost (MSB) flow chart.*

## 4 Experiments

### 4.1 Evaluation metrics

The primary metric and the training objective of the XGBoost model is the mean squared error of hourly data. However, in practice there may be alternative metrics that are more important to different customer scenarios. To understand how the predictions from ML models affect other metrics, the mean absolute error (MAE) of hourly data, RMSE of daily maximum and RMSE of daily minimum are also separately calculated. A daily maximum or minimum is defined as the maximum or minimum hourly temperature within a single day. It should be noted that this metric does not consider if the model can accurately predict the timing when the daily extreme temperature occurs. In addition, a critical error rate is defined as the number of hourly predictions with absolute error higher than 2.0 degrees, divided by total number of predictions made within validation period.

### 4.2 Results

Figures 5(a) and 5(b) demonstrate the evaluation metrics (hourly error, daily maximum and minimum error, and critical error rate as defined in 4.1) calculated on 6-month IMPROVER and MSB hourly data at a single site Hobart (27,

Oct 2022 to 27, Apr 2023). For hourly and daily extreme data, both RMSE and MAE are calculated and plotted with green and blue lines, respectively. Metrics are calculated on each forecast lead day separately. The results show that for this site, the employed ML model can reduce RMSE, MAE and critical error rate at all lead days. For RMSE of hourly data, the percentage of improvement on each lead day is similar, which shows that the ML model when trained on data with only lead day 0 can be applied on the prediction of all lead days. In most of the evaluated metrics and lead days for daily maximum and minimum predictions, the figures show that MSB has a better accuracy than IMPROVER. But exceptions also exist, such as the daily maximum dew point temperature on lead day 6 and 7. In addition, the results suggest that the improvements on RMSE and MAE have a similar trend, and therefore all subsequent discussions focus primarily on RMSE in this study.

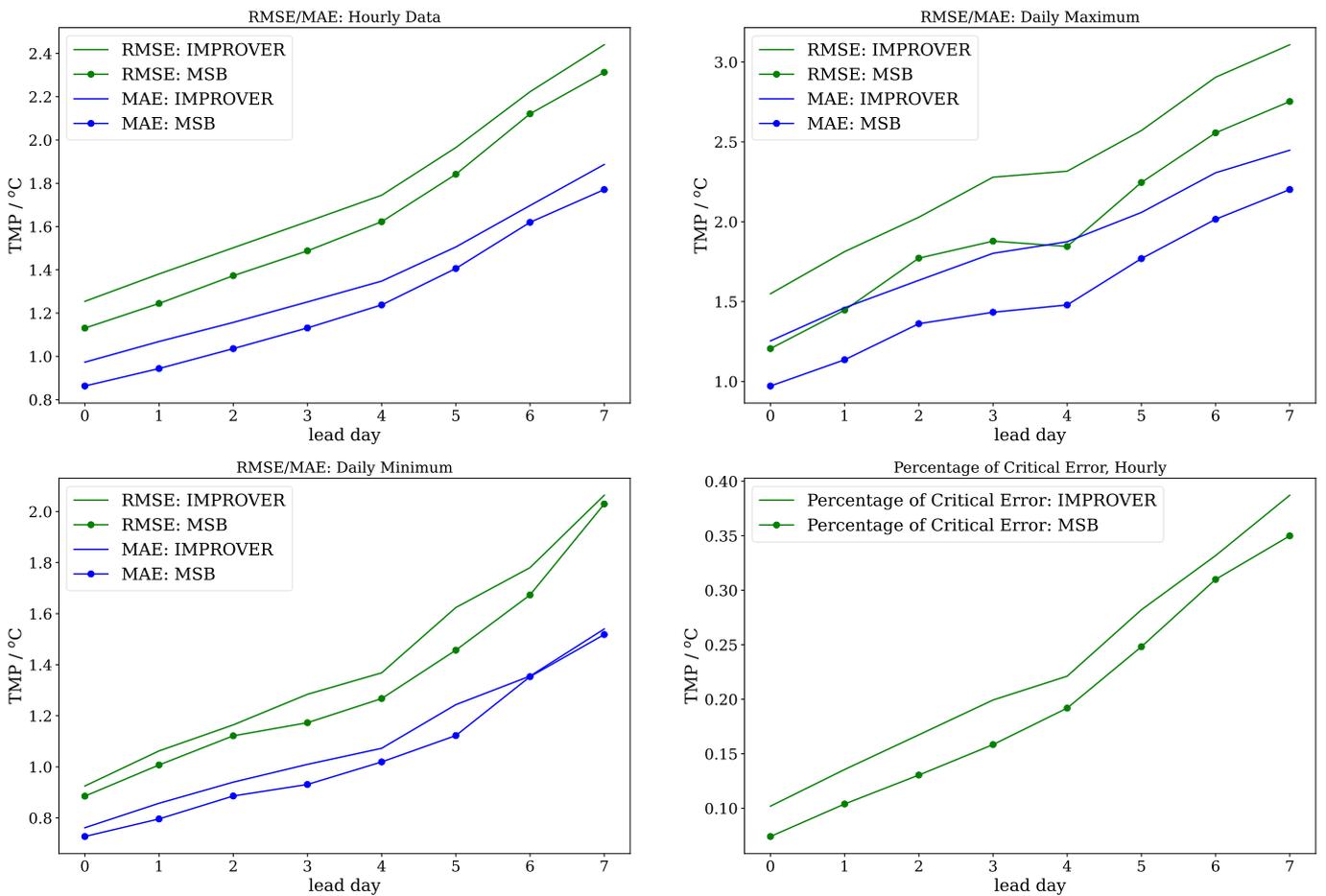

***Figure 5a***: *Comparison of Multi-SiteBoost results with IMPROVER grid values on various metrics. Site: Hobart (Ellerslie Road)*. *Temperature.*

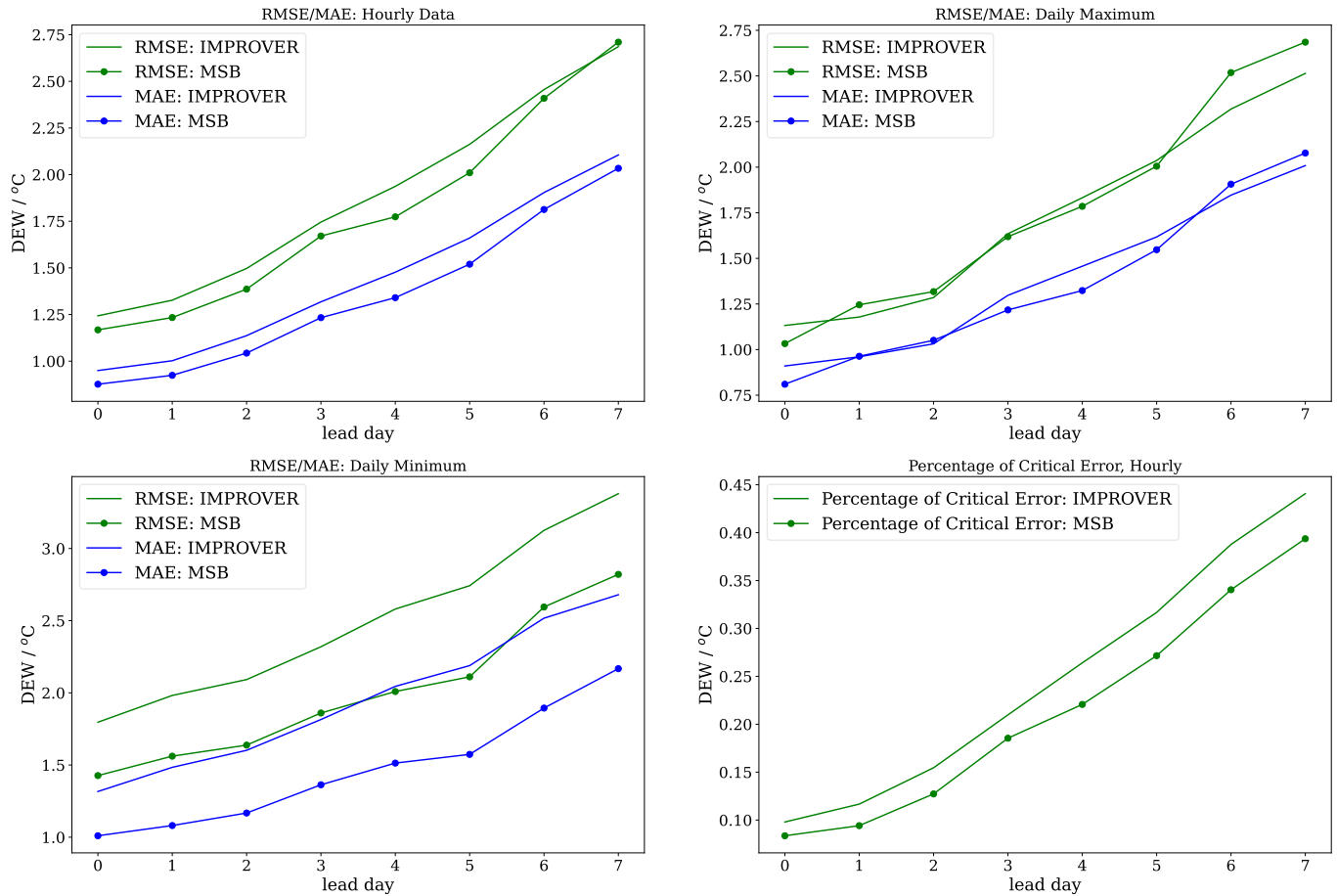

*Figure 5b:* Comparison of Multi-SiteBoost results with IMPROVER grid values on various metrics. Site: Hobart (Ellerslie Road). Temperature of dew point.

*Table 2:* Metrics of each site, calculated on all lead days, temperature forecast.

| **Site/Metrics** | Hourly RMSE, MSB (Change from IMPROVER) / ºC | Daily Maximum RMSE, MSB (Change from IMPROVER)/ºC | Daily Minimum RMSE, MSB (Change from IMPROVER)/ºC | Percentage of Critical Error, MSB (Change from IMPROVER)/% |
|---|---|---|---|---|
| Alice Springs Airport | 2.36 (-0.38) | 2.14 (-0.13) | 2.11 (-0.64) | 32.94% (-8.98%) |
| Archerfield Airport | 1.37 (-0.23) | 1.51 (-0.71) | 1.39 (-0.43) | 12.39% (-6.59%) |
| Avalon Airport | 1.92 (-0.07) | 2.11 (-0.14) | 1.77 (-0.34) | 23.50% (-1.31%) |
| Coffs Harbour Airport | 1.51 (-0.26) | 1.35 (-0.46) | 1.55 (-0.78) | 16.64% (-5.46%) |
| Curtin Aero | 1.79 (-0.13) | 1.67 (-0.33) | 1.33 (-0.24) | 20.60% (-2.73%) |
| Geraldton Airport | 2.09 (-0.60) | 2.22 (-0.68) | 2.09 (-0.87) | 29.67% (-13.45%) |
| Hobart (Ellerslie Road) | 1.64 (-0.12) | 1.95 (-0.37) | 1.34 (-0.06) | 19.53% (-3.30%) |
| Mount Isa | 2.19 (-0.21) | 1.96 (-0.25) | 2.00 (-0.56) | 31.95% (-2.63%) |
| Tindal RAAF | 1.82 (-0.16) | 1.57 (-0.14) | 1.38 (-0.15) | 22.70% (-5.34%) |
| Townsville | 1.13 (-0.20) | 0.99 (-0.60) | 1.15 (-0.04) | 7.44% (-4.39%) |
| Woomera Aerodrome | 1.93 (-0.11) | 1.83 (-0.04) | 1.75 (+0.07) | 21.44% (-4.28%) |

*Table 3: Metrics of each site, calculated on all lead days, dew point temperature forecast.*

| **Site/Metrics** | Hourly RMSE, MSB (Change from IMPROVER) / °C | Daily Maximum RMSE, MSB (Change from IMPROVER)/°C | Daily Minimum RMSE, MSB (Change from IMPROVER)/°C | Percentage of Critical Error, MSB (Change from IMPROVER)/% |
|---|---|---|---|---|
| Alice Springs Airport | 2.67 (-0.46) | 2.40 (-0.88) | 2.66 (-0.30) | 38.55% (-7.40%) |
| Archerfield Airport | 1.73 (-0.41) | 1.46 (+0.13) | 2.02 (-1.29) | 17.25% (-7.96%) |
| Avalon Airport | 1.64 (-0.01) | 1.51 (-0.35) | 1.75 (-0.18) | 19.00% (-1.65%) |
| Coffs Harbour Airport | 1.43 (-0.16) | 1.36 (-0.19) | 1.82 (-0.34) | 11.99% (-2.11%) |
| Curtin Aero | 1.99 (-0.02) | 1.69 (-0.10) | 2.14 (-0.12) | 23.63% (-0.85%) |
| Geraldton Airport | 2.39 (-1.05) | 1.63 (-2.26) | 2.47 (-0.27) | 31.16% (-22.45%) |
| Hobart (Ellerslie Road) | 1.80 (-0.09) | 1.79 (+0.05) | 2.01 (-0.49) | 21.38% (-3.47%) |
| Mount Isa | 2.68 (-0.53) | 2.38 (-0.12) | 2.75 (-1.16) | 38.16% (-5.78%) |
| Tindal RAAF | 1.55 (-0.12) | 1.27 (0.06) | 1.83 (-0.49) | 14.73% (-2.17%) |
| Townsville | 1.34 (-0.29) | 1.10 (-0.80) | 2.01 (-0.11) | 9.24% (-6.57%) |
| Woomera Aerodrome | 2.49 (-0.35) | 2.17 (-0.11) | 2.30 (-1.07) | 34.54% (-7.15%) |

*Table 4: Percentage of reduction of hourly RMSE by lead day, temperature forecast.*

| **Site/Lead Time (Days)** | 0 | 1 | 2 | 3 | 4 | 5 | 6 | 7 |
|---|---|---|---|---|---|---|---|---|
| Alice Springs Airport | 15.27% | 13.66% | 14.38% | 14.67% | 13.93% | 13.94% | 13.29% | 12.07% |
| Archerfield Airport | 16.10% | 14.87% | 15.45% | 13.19% | 13.71% | 13.72% | 13.58% | 14.21% |
| Avalon Airport | 3.75% | 3.32% | 1.75% | 2.81% | 6.91% | 5.60% | 4.63% | 5.67% |
| Coffs Harbour Airport | 16.22% | 16.30% | 16.45% | 15.10% | 13.70% | 14.71% | 14.44% | 14.91% |
| Curtin Aero | 1.46% | 1.79% | 3.54% | 5.36% | 6.55% | 8.82% | 10.19% | 13.06% |
| Geraldton Airport | 27.58% | 27.36% | 26.12% | 24.16% | 22.44% | 21.98% | 21.40% | 21.23% |
| Hobart (Ellerslie Road) | 9.86% | 9.88% | 8.59% | 8.32% | 7.01% | 6.29% | 4.59% | 5.22% |
| Mount Isa | 8.78% | 7.69% | 8.68% | 8.76% | 9.28% | 10.22% | 13.64% | 14.14% |
| Tindal RAAF | 7.18% | 6.25% | 6.13% | 8.78% | 9.36% | 7.09% | 8.19% | 8.50% |
| Townsville | 15.54% | 17.45% | 17.29% | 16.64% | 12.91% | 12.52% | 12.37% | 15.18% |
| Woomera Aerodrome | 4.34% | 4.71% | 6.64% | 7.76% | 6.73% | 9.33% | 5.80% | 5.86% |

*Table 5: Percentage of reduction of hourly RMSE by lead day, dew point temperature forecast.*

| Site/Lead Time (Days) | 0 | 1 | 2 | 3 | 4 | 5 | 6 | 7 |
|---|---|---|---|---|---|---|---|---|
| Alice Springs Airport | 10.70% | 11.74% | 15.30% | 17.89% | 18.30% | 14.66% | 12.89% | 13.30% |
| Archerfield Airport | 24.52% | 19.43% | 19.53% | 20.38% | 18.04% | 17.83% | 19.41% | 17.06% |
| Avalon Airport | 0.54% | 2.62% | 3.34% | 3.76% | 2.65% | 0.23% | -1.66% | -1.82% |
| Coffs Harbour Airport | 8.36% | 8.05% | 10.83% | 9.78% | 8.26% | 9.42% | 11.00% | 8.87% |
| Curtin Aero | -6.46% | -1.26% | 2.78% | 4.47% | 4.49% | 4.96% | 3.47% | 2.67% |
| Geraldton Airport | 37.13% | 33.37% | 31.56% | 30.58% | 30.59% | 30.50% | 29.48% | 28.14% |
| Hobart (Ellerslie Road) | 6.04% | 7.02% | 7.41% | 4.28% | 8.43% | 7.05% | 1.91% | -0.90% |
| Mount Isa | 4.62% | 8.64% | 14.55% | 18.51% | 18.44% | 16.60% | 19.67% | 21.49% |
| Tindal RAAF | 3.75% | 6.14% | 8.16% | 8.47% | 7.93% | 9.46% | 6.42% | 7.24% |
| Townsville | 10.61% | 11.57% | 14.16% | 15.77% | 22.43% | 24.06% | 19.77% | 14.35% |
| Woomera Aerodrome | 16.97% | 15.90% | 14.78% | 13.49% | 14.15% | 11.24% | 10.57% | 7.88% |

Tables 2-3 summarize the RMSE of hourly, daily maximum and daily minimum and percentage of critical error calculated for each site from all lead days. The change of each metrics from IMPROVER is also included in the tables, and a negative change means a lower RMSE or critical error percentage of MSB results compared with IMPROVER. The results show that MSB reduces hourly RMSE on all sites from up to 1.05ºC, but effect largely varies with site location. In general, the sites where forecast error is already low in IMPROVER (e.g. Avalon Airport) benefits less from MSB than those with originally higher forecast error (e.g. Geraldton Airport). In addition, as shown in Figures 5(a) and 5(b), the reduction in forecast error largely varies with forecast lead day, so the actual reduction on each lead day may be significantly different from the averaged reduction.

Tables 4-5 summarize the improvement of prediction accuracy on all sites and all lead days, in terms of percentage of reduction of RMSE from MSB as compared to IMPROVER predictions. In this way, a positive percentage in the tables indicate a lower RMSE on MSB data and an improvement compared with IMPROVER. For sake of brevity, other metrics as discussed in section 4.1 are listed in Appendix. As shown in the tables, most of the calculated metrics are positive, which indicate a general forecast improvement on tested sites due to the applied ML models. However, the tables also show variation of ML skills over the selected sites. Even on hourly RMSE on lead day 0, which is the training objective, the percentage of RMSE reduction can range from 0.54% (Avalon airport, dew point temperature) to 37.13% (Geraldton airport, dew point temperature). And in Curtin Aero, the MSB predictions are worse than the input IMPROVER data, as the RMSE of MSB predictions is 6.46% higher than that of IMPROVER. Averaging on all sites and all lead days, MSB has an average reduction of hourly RMSE by 11.35% for temperature and 12.28% for dew point. Critical error is reduced by 5.60% for temperature and 6.19% for dew point. The improvement in RMSE and the percentage of reduction in temperature is comparable to post-processing studies using non-ML methods such as Delle Monache et al. (2011) and Sheridan et al. (2018).

The variation of ML skills on various site locations highlights on tested sites highlights the importance of localised model training and tuning. In the data pre-processing, most of the parameters are site-agnostic and optimized based on averaged performance over all selected sites. All sites share the same pool of available features. This leads to less training time, a simpler data pipeline and ensures a good overall performance. But it is not guaranteed that the model is tuned to its maximum performance on each site. When customer needs arise for the optimization of some certain site, further experiments on input features and parameters on specific sites of concern should be performed.

Figure 6 shows the histograms of pointwise forecast error distribution of IMPROVER and MSB for a single site on lead day 0. The results show that error distribution of MSB data is more zero-centred and has a narrower bandwidth, which implies both a smaller mean error and less critical error rate of the dataset. However, the figure also shows that MSB error distribution is not perfectly Gaussian, which implies that systematic error still exists even though the overall performance is already improved compared with IMPROVER grid data.

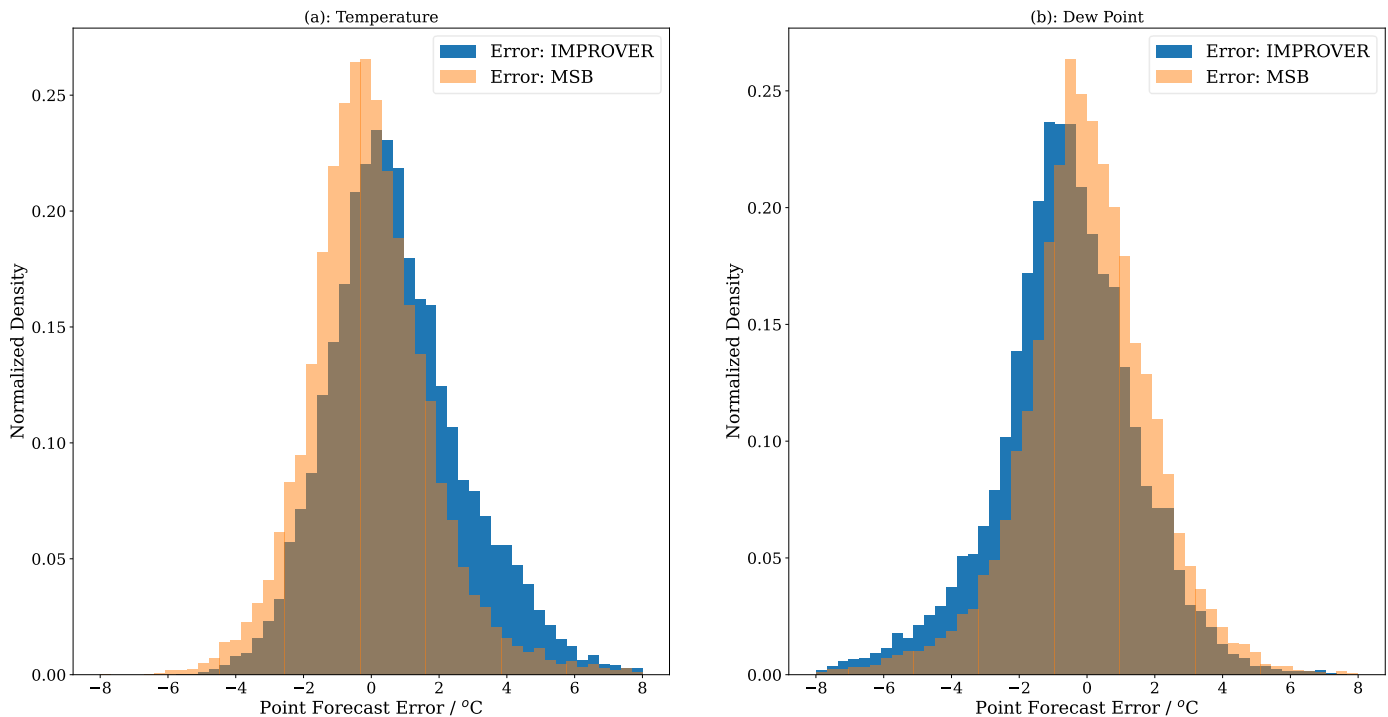

*Figure* 6: *Histograms of pointwise forecast error on lead day 0 for (a) Temperature; (b) Dew point temperature. Site: Hobart (Ellerslie Road).*

## 5    Explaining model outputs with SHAP

Figure 7(a) shows the SHAP summary plot on the scaled training set of a temperature forecast model. In the summary plot, X axis is the SHAP value, and the 10 features used by this model are arranged along the Y axis. On each of the rows, every point represents a sample in the training set. The points are coloured by feature value. The rows of features are arranged by relative magnitude of SHAP values in descending order. The summary plot shows that for this model, the four selected temperature measurements $T_{W}$, $T_{SE}$, $T_{S}$ and $T_{SW}$ have the largest overall contributions, indicated by the extent of their SHAP values. A physical explanation for the degree of contribution of $T_{W}$, $T_{SE}$, $T_{S}$ and $T_{SW}$ has not been explored. This agrees with the intuition that the actual temperature is most relevant to the predictions from NWP models. Other supplementary features such as hour, wind and dew point at various locations in general have lower SHAP value magnitude, which implies they give minor changes to each prediction. However, the rows of $T_{dSE}$ and $T_{dS}$ also show at certain samples, a high dew point can largely decrease the final predicted values of temperature. But this effect is asymmetrical, as a low dew point doesn't result in a large positive SHAP value.

Figure 7(b) is the global feature importance plot, calculated by averaging all SHAP value magnitude of each sample for each feature. The resultant feature importance for this model agrees with XGBoost feature importance, calculated by maximum gain. Compared with Fig 7(a), Fig 7(b) lacks the information of the distribution of feature importance on each sample. For example, it doesn't show the occasional significance of $T_{dSE}$ on some certain points. This comparison highlights the importance of local explanation provided by SHAP, as supplementary information in addition to global feature importance.

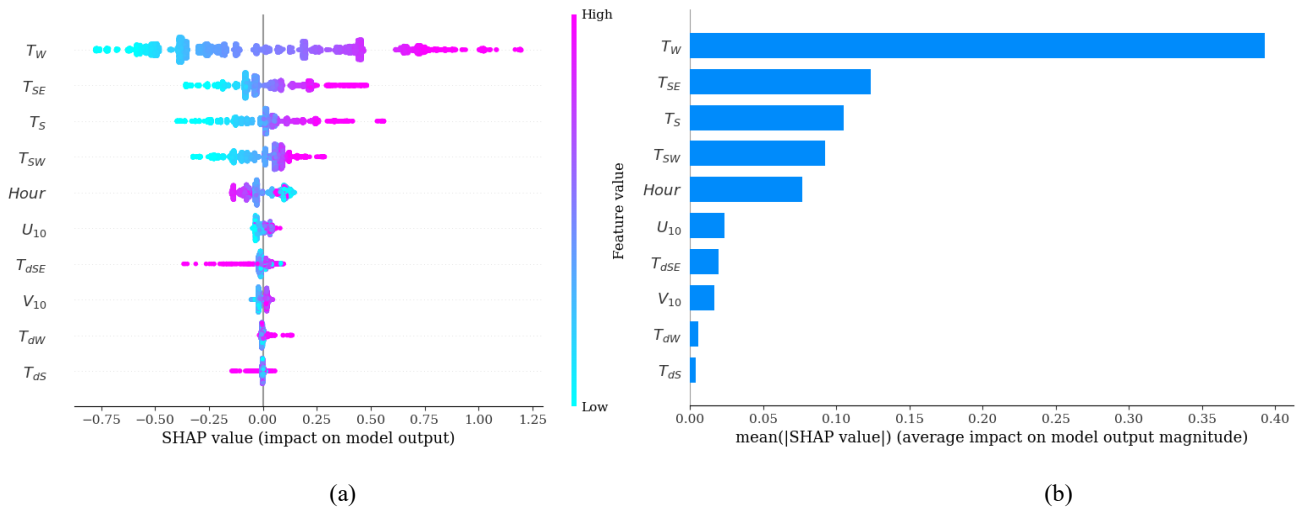

*Figure 7:* Global view of SHAP values on training set for top 10 features, (a) SHAP summary plot, coloured by feature value of each point; (b) Feature importance, calculated by aggregation of absolute value of SHAP of each point. Training window: 2023-01-18 – 2023-3-22; Trained variable: temperature; Site: Archerfield Airport.

Figure 8 plots the SHAP values of all samples with respect to feature values for three features: $T_w$, *Hour* (in UTC) and $U_{10}$ in a model trained for temperature prediction. The plots show that the SHAP value of feature $T_W$ is approximately linear with the feature value. The vertical dispersion is minor, and each value of $T_W$ approximately corresponds to one single SHAP value. This agrees with Figure 2(a), both of which can be interpreted as the NWP grid value is linearly correlated to site observation. By comparison, the SHAP value of hour (defined as a discrete integer rather than a continuous variable) is not linearly correlated to the feature value, and one value of hour can correspond to multiple SHAP values. However, it can still be observed that hour 0-5 generally has positive SHAP while hour 15-20 generally negative. This agrees with observations and statistics, as temperature on this site at UTC hour 0:00 – 5:00 (10:00 – 15:00 local time) and 15:00-20:00 are expected to be hotter and colder than daily average, respectively. The SHAP values of wind by comparison are highly nonlinear, multi-valued and difficult to interpret, which indicates a highly nonlinear relationship between wind speed and temperature, and strong interactions between wind speed and other features.

The interactions between features can be more clearly visualized by SHAP interaction value matrix, as shown in the heatmap Figure 9. In Figure 9, the diagonal terms are the main effect while other terms are interaction effects between two features. The cells are coloured by the magnitude of the cell values. To better show the relative magnitude of interaction effects compared with main effects for each feature, each column in the plot is scaled separately so that the diagonal terms (the main effects of each feature) are scaled to 1.0. Consequently, the matrix is no longer symmetrical. The matrix shows that the interaction effects of four temperature features $T_W$, $T_{SE}$, $T_S$ and $T_{SW}$ with other features are much smaller than their main effects. This agrees with Figure 7(a) in that the SHAP values of temperature depends almost only on the feature value. By comparison, the interaction effects are more significant for dew point and wind speed, as some of the interaction effect terms are of similar magnitude with the main effects.

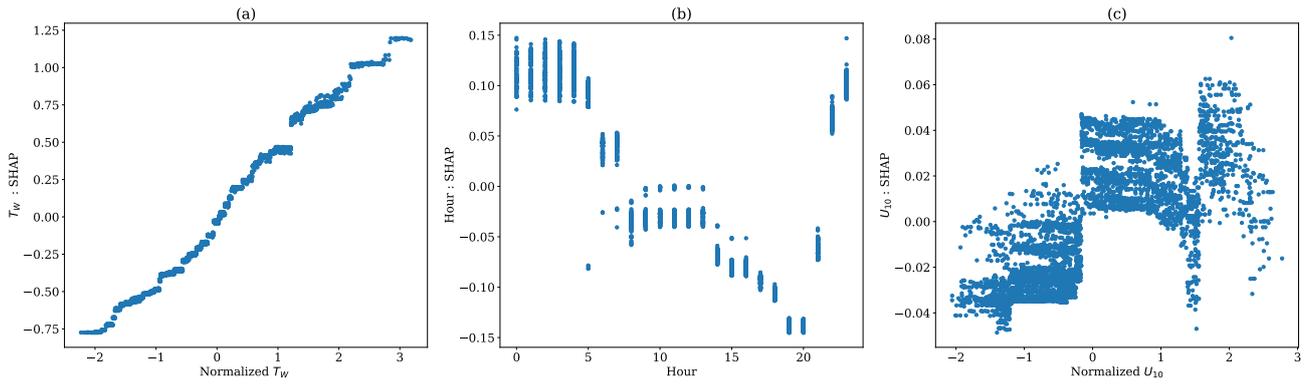

**Figure 8:** *SHAP values of three features for all sample points in training set. Training window: 2023-01-18 – 2023-3-22; Trained variable: temperature; Site: Archerfield Airport.*

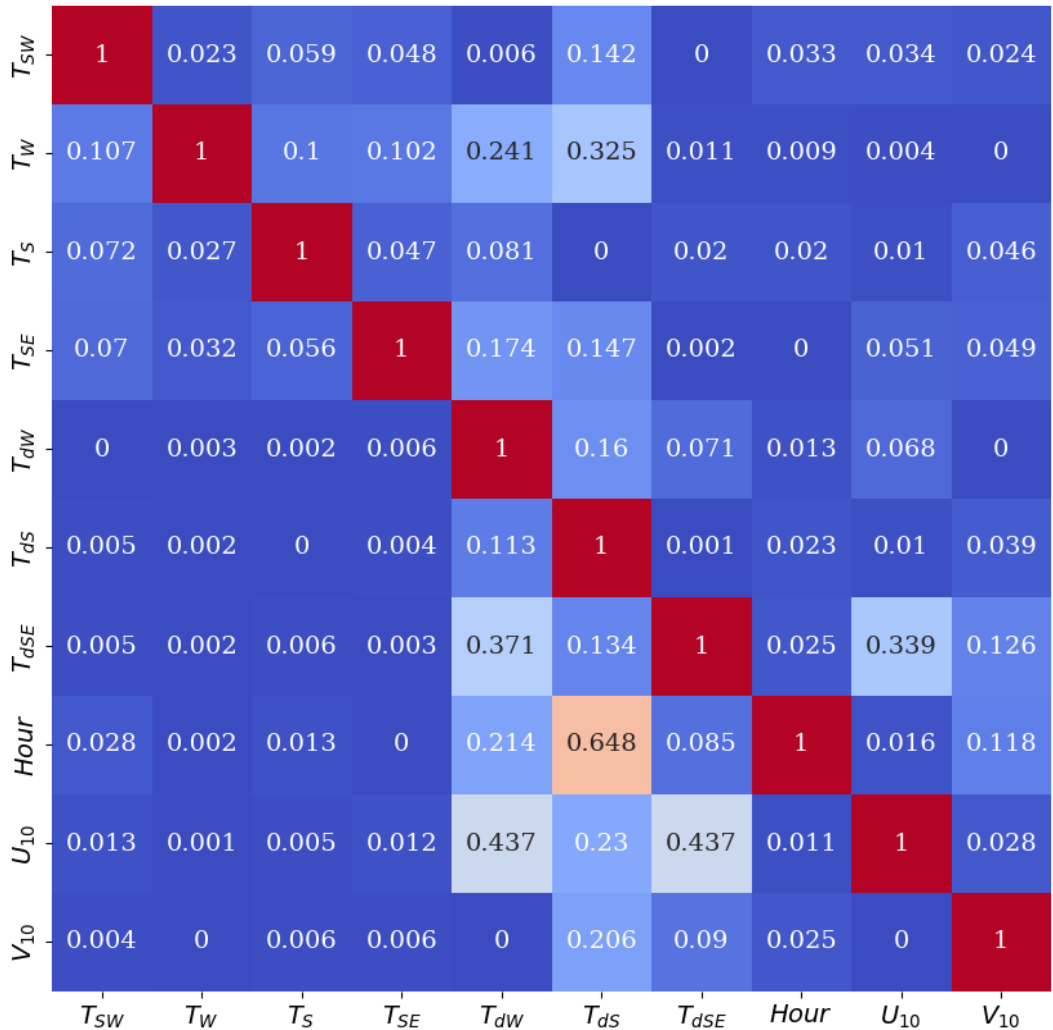

**Figure 9:** *Normalized SHAP interaction values between each pair of features (normalized at each column), Training window: 2023-01-18 – 2023-3-22; Trained variable: temperature. Site: Archerfield Airport.*

The analysis of the relationship between each feature and its corresponding SHAP value reveals some insights into the mechanism of the temperature forecast model. The model output can be approximately described as a combination of a main component from linear combinations of selected NWP predictions of the main variable at different locations of the 3x3 grid, and some supplementary components of lesser magnitudes. The supplementary components are related to features that are other than the predicted variable, such as hour, dew point, wind speed etc. The effects of these features are highly nonlinear but can be significant to the model predictions at some points. These two categories of features are hereinafter referred to as 'linear features' and 'nonlinear features', respectively. The dew point forecast models are found to exhibit similar patterns as the temperature forecast model, so that the discussions on them are not repeated here.

As discussed in Introduction, SHAP values are used for explaining and understanding model outputs, but do not include any information on the accuracy of the model. However, the insights can be used to debug and monitor the model behaviour. The next section will focus on analysing model error to increase ML reliability, in which SHAP values play an important role.

## 6  Error analysis and model reliability

ML models are imperfect and often contain systematic errors, even if the level of accuracy is satisfactory when evaluated based on given error metrics. Recall that the error plot in Figure 6 shows that the distribution of MSB error is not perfectly Gaussian. Since the trained model generates single predictions without giving information on confidence interval, it is possible that the model seems over-confident even when the prediction is likely to be inaccurate. Therefore, it's important to analyse and understand model error so that model accuracy can be consistently monitored and improved. In this section, we aim to apply several error analysis approaches to MSB framework, so that individual predictions that are considered unreliable and may be subject to higher error than average can be discovered at the same time the predictions are made. Hereinafter an 'unreliable prediction' refers to an individual prediction point with prediction error higher than overall metrics. As a benefit, when a weather forecast from MSB is potentially unreliable, customers will be notified and advised to rely on alternative sources of information for these points.

This section presents the various methods incorporated into MSB to discover unreliable and potentially less accurate predictions before the observation data is made available, so that alternative information source can be sought in advance as a supplement to MSB predictions. The reliability study is only applied to forecast with lead day 0 because forecast with longer lead days can be subject to change with time as new observation becomes available. Note that since our goal is to provide foresight to the predictions, we do not apply posterior analysis approaches that require the knowledge of ground truth in this study. Meanwhile, we aim at keeping the process fully automatic and time efficient, so the adopted approaches are selected with the consideration of their simplicity together with their effectiveness. Meanwhile, it should be noted that studies in this section aim at identifying high-error predictions only, without separating the source of error into aleatoric and epistemic uncertainty.

### 6.1  Unreliable predictions from out-of-bound feature values

As already discussed in Section 3, XGBoost models do not extrapolate, which means if a feature value is not within the range of the training set, the effect of the feature may not be accurately accounted for by XGBoost. Data scaling is applied during pre-processing to counter this problem as previously described in Section 5. The method has proved highly effective, but does not fully eliminate the presence of out-of-bound feature values in the validation set. Especially when a high-importance feature has an out-of-bound value, the model prediction is more likely to be less accurate than average.

This error source can be best visualized by plotting feature value with respect to its SHAP value. Consistent with the convention in Figure 8, Figure 10(a) plots the scaled feature values vs the corresponding SHAP values of the most important feature, $T_{\text{dSW}}$, in a dew point prediction model. The model is trained on data from 2023-01-18 to 2023-03-23 and validated on the 2023-03-24 to 2023-03-31 for site Archerfield Airport. All sample points in the training set are plotted with blue circles while those on validation set with green squares. All points are coloured by pointwise absolute error. Similar to the exemplar model explained in Section 5, the SHAP value of $T_{\text{dSW}}$ is approximately linearly correlated to its feature value. In the validation set, the samples whose feature values are within the range of training set (around -2.0 ºC to 2.5 ºC) largely overlap with the samples in the training set. However, for validation samples with out-of-bound feature values (highlighted by red square region), the SHAP values no longer increase with the feature values, which means the effect of these feature values are likely to be underestimated by the model. Correspondingly, these points are coloured by dark green, indicating much higher error than rest of the samples. This results in the high-error predictions in the time series plotted in Figure 10(b), in which the high error points within the red square in Figure 10(a) are highlighted in red. The figure shows that a sudden drop in dew point occurred on 2023-03-30 due to very low values that were not experienced in the entire training period. As a result, the trained model significantly overestimates the dew point temperature (as did IMPROVER).

Based on the analysis above, we include in MSB an automatic detection of potentially unreliable predictions by examining if the value of the most important feature of any validation point is outside the range of the training set. Figure 11 plots the RMSE and critical error rate of recognized unreliable points as compared to the metrics on all data points for each site. Only predictions with lead day 0 are considered. The results show that for almost all sites, the unreliable subset has a higher error than global. For some sites, the RMSE of unreliable subset can be twice as high as average (site Curtin, dew point temperature).

It should be noted that the appearance of this type of error is very rare. The selected subset composes around 0.5% of all data. However, this error analysis does highlight the importance of data scaling, without which more out-of-range feature values would appear in the validation set.

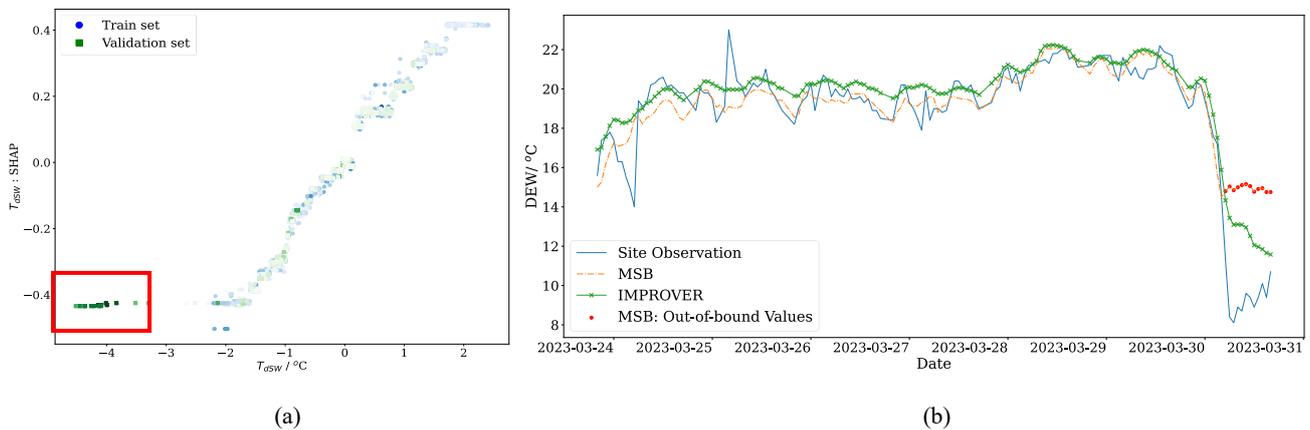

*Figure 10:* Example of inaccurate outputs from XGBoost due to out-of-bound feature values, (a) feature values vs. corresponding SHAP values for data samples in training set and validation set, coloured by absolute error of the prediction; (b) Time series of XGBoost predictions in the validation period and site observations. Inaccurate predictions due to out-of-bound feature values are highlighted in red.

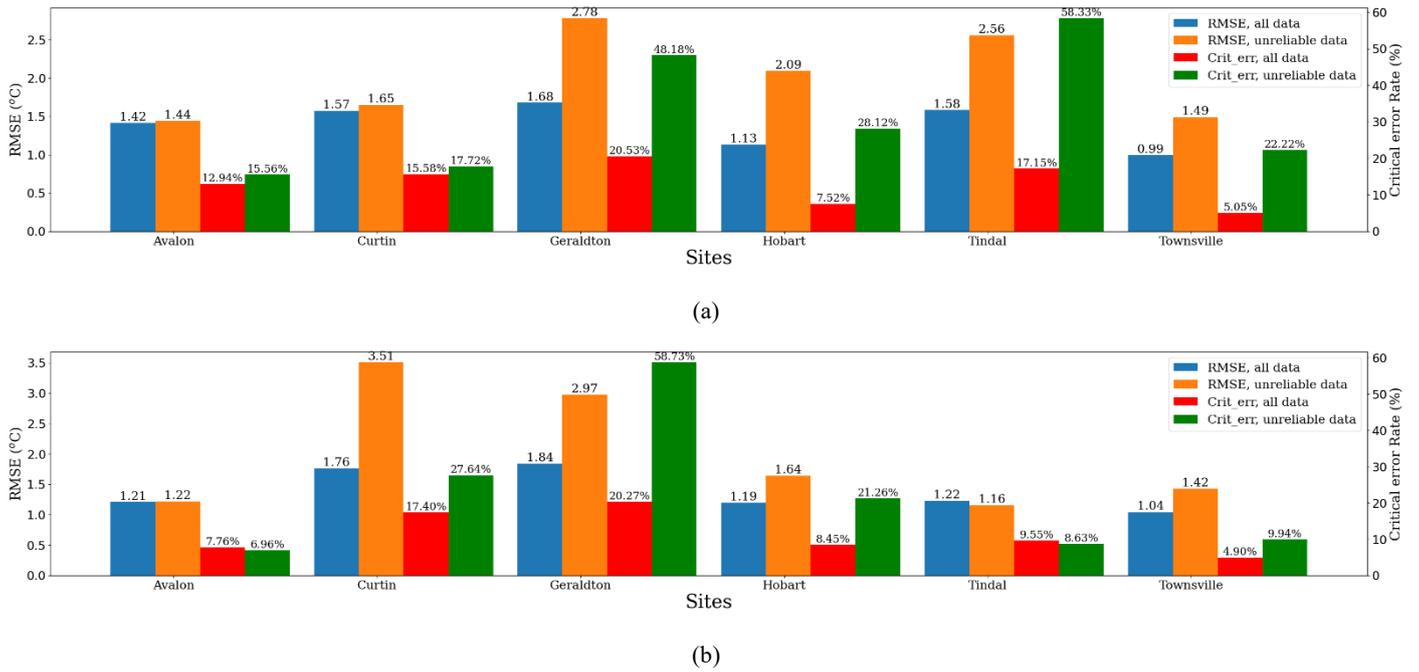

*Figure 11:* Statistics of RMSE and critical error rate of all predictions and unreliable predictions due to out-of-bound feature values. (a) Temperature; (b) Dew point temperature. Lead day: 0.

## 6.2 Unreliable predictions in poor-performing data cohorts

It is generally understood that error often isn't evenly distributed among the entire dataset, and a subset (or equivalently, a "cohort") of data could have worse accuracy than others. However, the underperforming data cohorts are sometimes difficult to find, and various approaches have been proposed in literature to discover them as summarized by (d'Eon et al. 2022). In this study, we utilize the method adopted in the Python package *InterpretML* (Nori et al. 2019), which involves training a single decision tree on the dataset, with scaled feature values of the data samples as features, and prediction error as labels. Figure 12 demonstrates an example of such decision tree, trained with data samples from the same model analysed in Section 6.1 and visualized by *InterpretML*. The value in each node represents the MAE of the data that lies within the node, and the filling level of each node represents the proportion of errors associated with that node, coloured by magnitude of error. The figure shows that while the MAE of all data samples is 0.67, the cohort of data with feature values $-0.31<U_{10}<=-0.16$ and *Hour* $<=7.50$ is subject to a higher MAE of 0.90. This cohort comprises 13.5% of all data, but contributes to 18.22% of all error. The example shows that a single decision tree can effectively cluster the data and divide it into high and poor performing cohorts.

Assuming no significant model and data drift in the validation set, the data and predictions in the validation set should have similar statistics as in the training set. Therefore, we include a second unreliable prediction identification in MSB, by training a single decision tree on the prediction error in the training set, and identify high error data cohort and its corresponding feature values. When data samples in validation are identified to lie within the high error cohort by their feature values, the predictions from these samples are considered unreliable. A high error data cohort is defined as having RMSE twice as high as the global RMSE in the training set, and the minimum amount of data it covers is 1.0%. These thresholds are empirical and defined based on parametric studies.

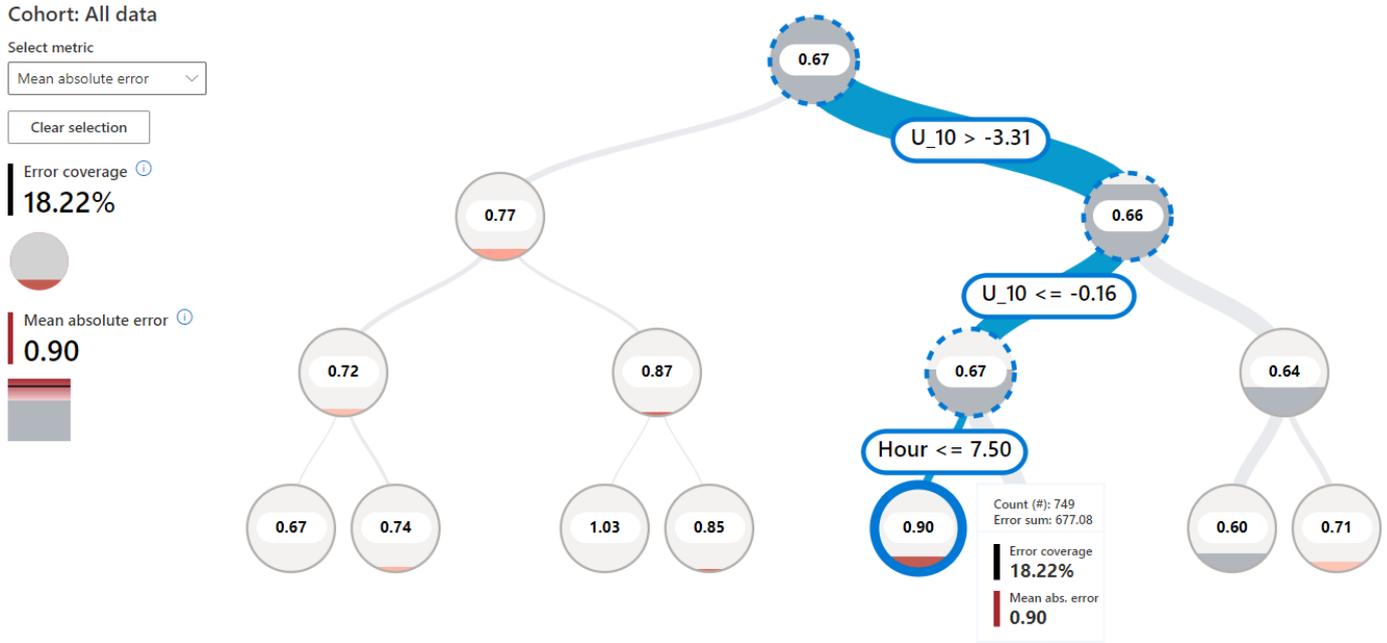

*Figure 12: High error data cohort identified by a single decision tree on training set, visualized by InterpretML ErrorAnalysisBoard. Training window: 2023-01-18 – 2023-3-22; Trained variable: dew point temperature. Site: Archerfield Airport.*

Figure 13 demonstrates the RMSE and critical error rate of all predictions and unreliable predictions identified with this approach. The identified data comprises 7.3% of all prediction points on average. And the results shown in Figure 14 indicate that the method is also widely effective among various sites. The RMSE of unreliable predictions are 1.1-1.7 times greater than average.

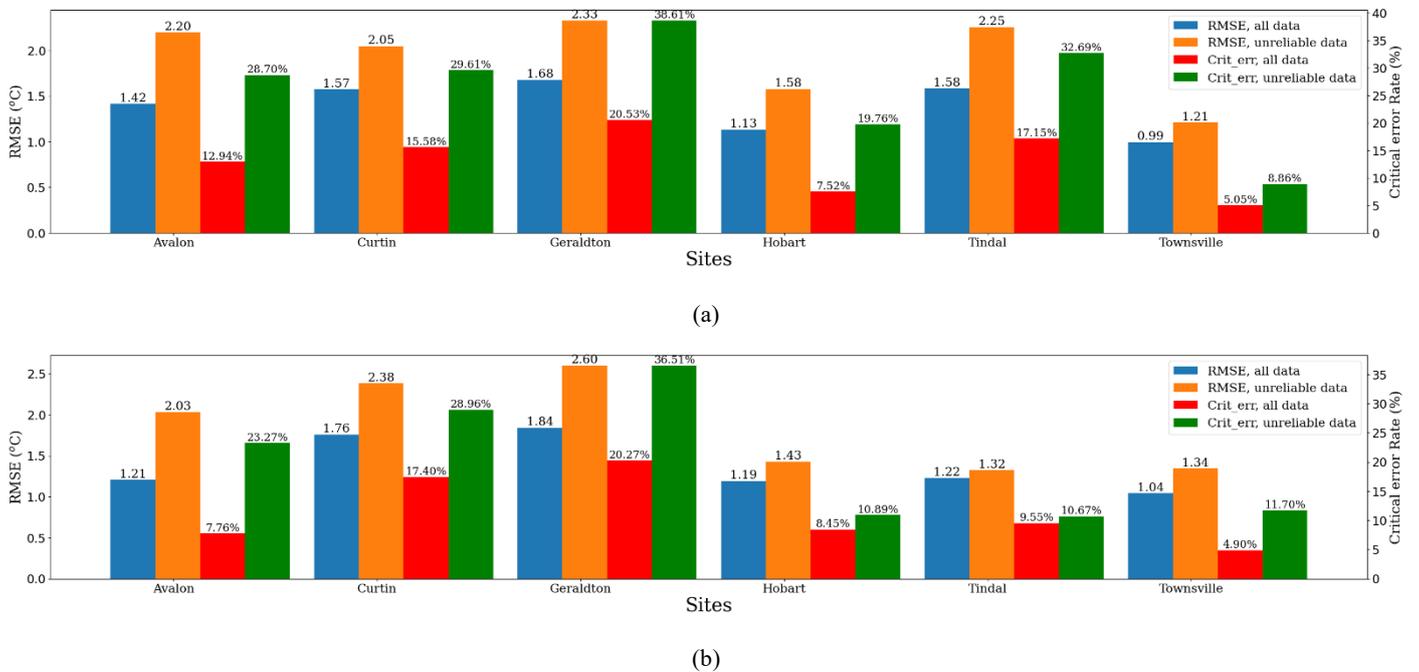

*Figure 13: Statistics of RMSE and critical error rate of all predictions and unreliable predictions in high error data cohorts. (a) Temperature; (b) Dew point temperature. Lead day: 0.*

## 6.3 Unreliable predictions without local fit

To improve the reliability of a ML model, Saria and Subbaswamy (2019) proposed two principles to determine reliability of a single sample point: (1) density principle: the tested sample point should be close to the training set; (2) local fit principle: the model should predict accurately on the training points closest to the tested sample point. Based on these two principles, a prediction of the ML model can be considered reliable when the tested sample is close to the training set and lies in the region where the model prediction is accurate.

The density principle has been taken into consideration in Section 6.1, in which the distance of a sample point to the training set is measured by its main linear feature only. Subsequent experiments have shown that alternative approaches to account for the density rule do not perform better in identifying unreliable predictions. The local fit principle is partially considered in Section 6.2, in which samples with high prediction error are sorted out by a single decision tree. During the training of the decision tree, however, we must ensure each leaf contains enough number of samples, so very rare cases may not be separately labelled during the process. Consequently, we add an alternative approach here by identifying all test sample points whose closest points in training set are predicted poorly by the model. The closest 3 points in the training set are found by nearest neighbour algorithm (KNN) (Hellman 1970). And local accuracy of the 3 points is determined by their average absolute error. If the average error is as high as 90-percentile of all training error multiplied by 2.0, the local accuracy is considered low, and the prediction of the sample point is considered unreliable.

The high error threshold as described above is set empirically and higher threshold results in fewer points identified as being unreliable. Figure 14 demonstrates the RMSE and critical error rate of all predictions and unreliable predictions identified with this approach. The identified data comprises of all prediction points on average. And the results shown in Figure 14 indicate that the method is also widely effective among various sites. The RMSE of unreliable predictions are 0.6-1.5 times greater than average.

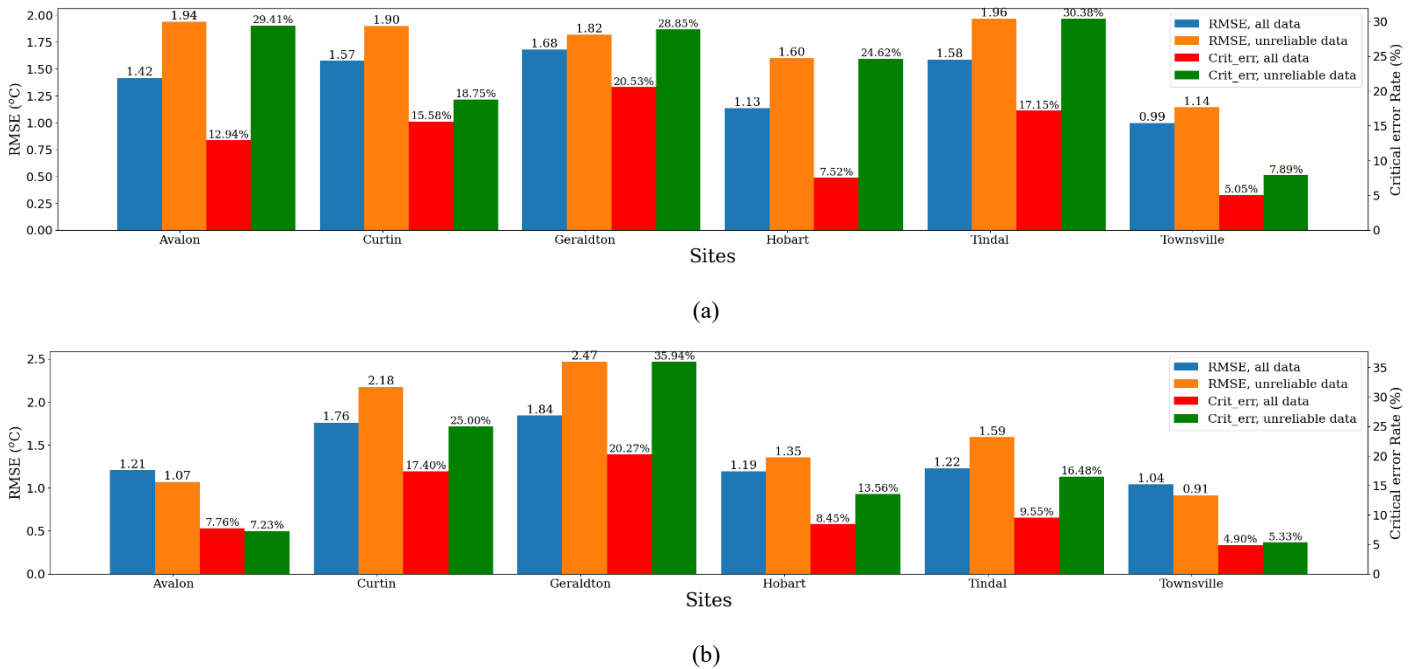

*Figure 14:* Statistics of RMSE and critical error rate of all predictions and unreliable predictions without local fit. (a) Temperature; (b) Dew point temperature. Lead day: 0.

## 6.4 Summary of Error Analysis and Model Reliability

In this section, different methods of unreliable prediction identification that are incorporated in MSB framework are introduced. When evaluated separately, all these methods show satisfactory results on various sites, since the data identified as unreliable indeed has significantly higher error in general. These methods approach the same problem with different aspects, so their results are not exclusive from each other. One single prediction can be identified as unreliable by more than one method. Note that the methods applied here are not exhaustive, and alternative methods and software packages exist in the literature. The methods and the empirical parameters are selected here based on their effectiveness and simplicity, so that unreliability identification process does not significantly influence the efficiency of the framework.

Figure 15 shows the accuracy of unreliable predictions identified by all methods combined and its comparison to global accuracy. The results show that the overall performance of the methods is satisfactory, with the RMSE of unreliable samples 1.1 - 1.5 times as high as site average. The contributions of each separate method have been demonstrated in previous sections.

Figures 16a and 16b show the time series of IMPROVER and MSB predictions for temperature and dew point in January 2023 and their comparison with observed values. Identified unreliable predictions are highlighted as red dots along the MSB curve. Some predictions with large prediction errors have been successfully identified, such as temperature predictions at 5th Jan and 19th Jan, and dew point predictions at 5th Jan and 15th Jan. The identification of unreliable points agrees with intuition, since predictions on 5th Jan are visually unstable and oscillate with time. These low-cost identification methods do not aim to accurately capture all high-error predictions, so there are misses (e.g. temperature on 31st, Jan) and false-alarms (e.g. temperature on 12[th], Jan). In addition, the identification gives no information on whether the unreliable predictions are less accurate than IMPROVER input, For example, the minimum temperature prediction on 15[th], Jan in Figure 16(a) is marked as unreliable, but it is still closer to the observed value than IMPROVER.

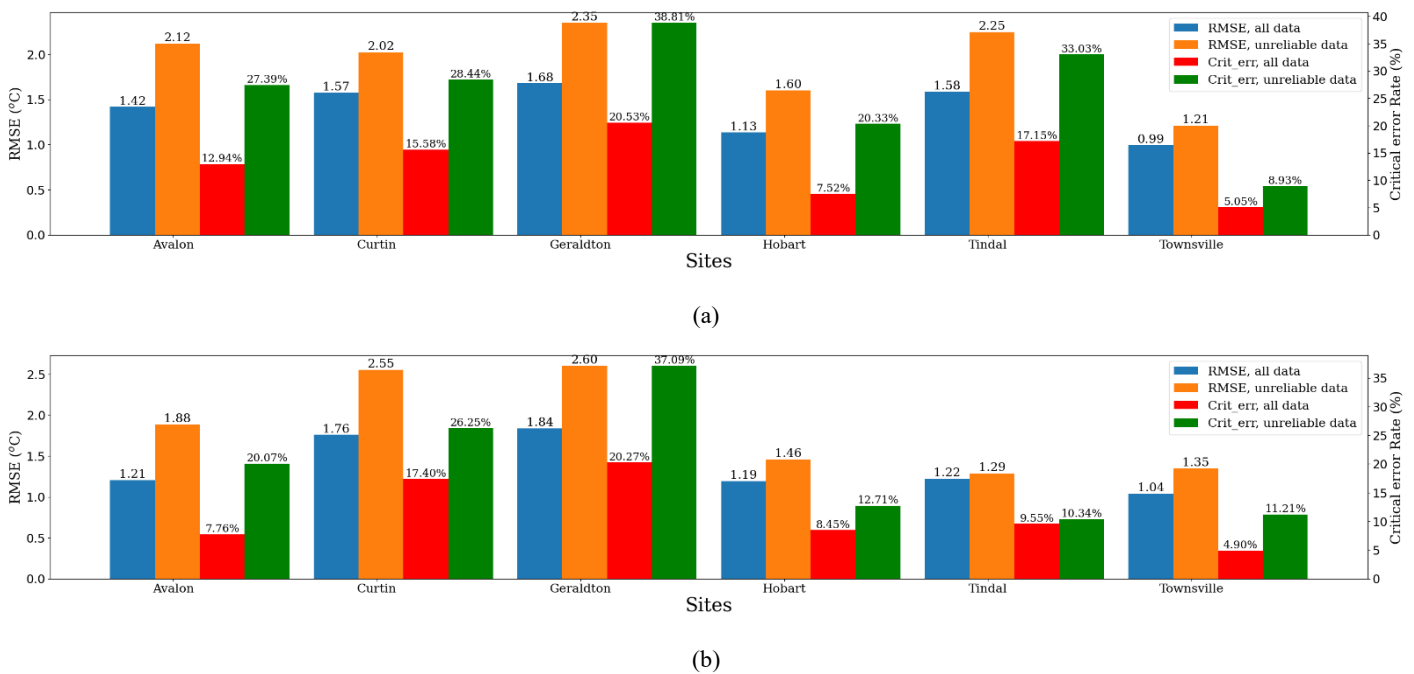

*Figure 15:* Summary of RMSE and critical error rate of all predictions and unreliable predictions due to all possible error types. (a) Temperature; (b) Dew point temperature. Lead day: 0.

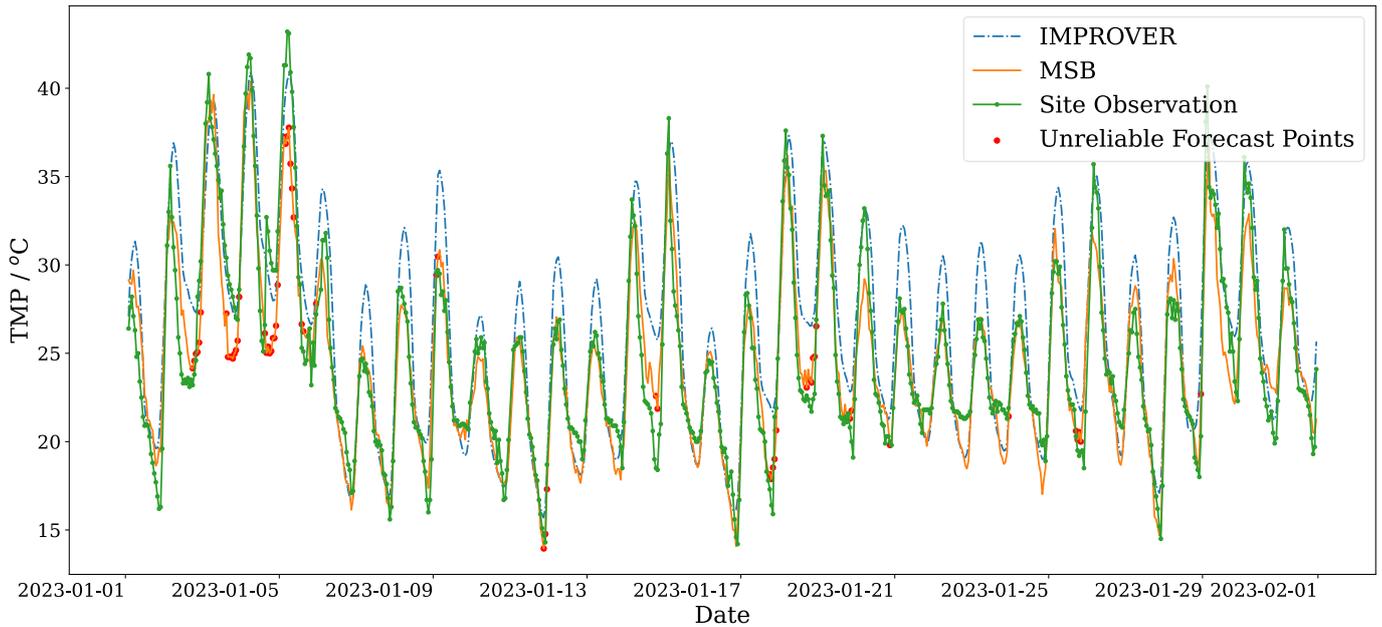

*Figure 16a: Exemplar time series of XGBoost prediction, compared with IMPROVER grid value and site observations, unreliable data highlighted by red. Lead time: 0 – 24 hrs; Site: Geraldton. Temperature.*

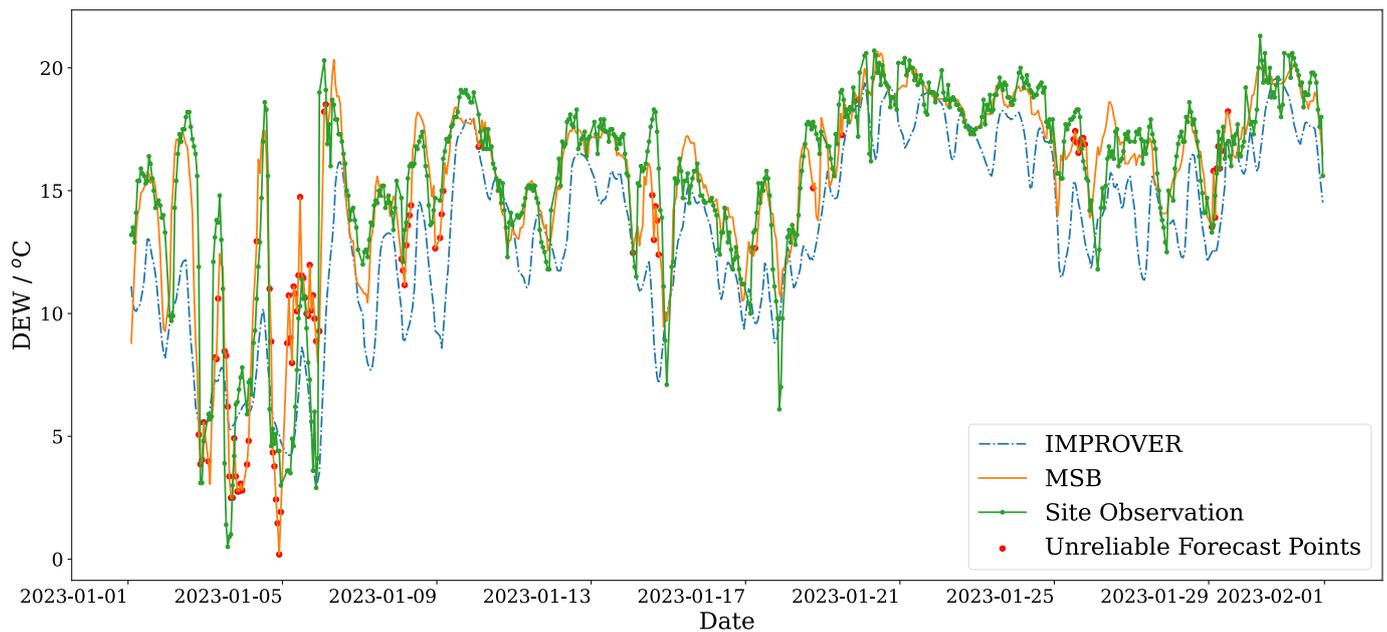

*Figure 16b: Exemplar time series of XGBoost prediction, compared with IMPROVER grid value and site observations, unreliable data highlighted by red. Lead time: 0 – 24 hrs; Site: Geraldton. Dew point temperature.*

# 7  Conclusions and future work

This study focuses on the proof-of-concept of applying XGBoost, a state-of-the-art ML package for the application on tabular data, to hourly site temperature and dew point forecasts up to 7 days ahead, showing test results for various sites located across Australia. The resultant ML framework, Multi-SiteBoost, comprises the following 3 significant aspects: (1) an optimized data pre-processing pipeline, with each data transformation technique proven effective based on parametric study; (2) the application of SHAP values for model explainability, and (3) the identification of individual unreliable predictions at the moment when the prediction is made, so that extra caution can be taken for time spots when predictions are unreliable and may be subject to higher error.

The predictions generated by the proposed ML framework are evaluated on both RMSE and various customized metrics, and satisfactory prediction accuracy has been achieved on most of the metrics at various sites across Australia at various lead days. Across the selected sites, an overall reduction in RMSE of 11% for temperature and 12% for dew point is estimated on average, as compared to grid values of IMPROVER, (which itself is already blended and post-processed from multiple raw NWP data sources). By improving the general forecast accuracy, improvement on other related metrics such as RMSE of daily minimum and maximum, and critical error rate, have been improved as well - even if the ML models are not trained with those objective functions. However, model accuracy shows noticeable variance across sites, highlighting the necessity of local parameters and hyperparameter tuning instead of a global setup. This is enabled in the current framework, granting flexibility in case of customer interest in the future.

Explaining the trained ML models with SHAP values and SHAP interactive values leads to additional insights and trustworthiness of the model. The explanation identified the different effects of each variable, which largely agree with both global explanation of feature importance and human experience. The identification of main 'linear' features also confirms the significance of proper data scaling prior to training and paves the way for the identification of some unreliable predictions.

To add additional information to the predictions and as a forecast-auditing procedure, pointwise reliability is evaluated with three different approaches. The unreliability due to out-of-bound feature values is deduced by the inherent property of XGBoost algorithm, while unreliability identified by high error cohorts and local unfit rule are statistical methods based on the model accuracy on the training set. These approaches combined can identify around 5% - 10% of all predictions as unreliable and these data have RMSE of up to 1.5 times higher than global RMSE on tested sites. The pointwise reliability provides a foresight on local prediction quality prior to the observation data becoming available. As an auditing process of the forecast, it adds additional information of confidence level to the deterministic forecast. Like the accuracy of the ML model, the effectiveness of these methods also varies with site locations.

It should be noted that while the current study explores specific data ingests, the developed methodology has general application, as the ML training, validation and explanation in this paper does not rely on specific data properties. However, higher quality data and relevant features certainly improve the prediction quality given the same ML architecture. Future work includes benchmark studies on similar datasets to determine the effectiveness of XGBoost compared with conventional statistical techniques, fine-tuning on customer-specified sites with higher significance, generating probabilistic forecasts and adding nowcast capabilities to the existing MSB framework, so that real-time observation data can be ingested to correct forecasts in short forecast windows (1 to 6 hours ahead).

## Conflict of interest

No potential conflict of interest was reported by the authors.

# Appendix A: Supplementary Data

*Table A.1:* *Percentage of reduction of daily maximum RMSE by lead day, temperature forecast.*

| Sites/ Lead Time (Days) | 0 | 1 | 2 | 3 | 4 | 5 | 6 | 7 |
|---|---|---|---|---|---|---|---|---|
| Alice Springs Airport | -14.36% | -8.69% | 4.21% | 9.33% | 9.74% | 14.12% | 10.22% | 5.89% |
| Archerfield Airport | 33.62% | 33.97% | 33.27% | 29.04% | 29.46% | 29.15% | 32.26% | 35.17% |
| Avalon Airport | 10.93% | 15.30% | 3.91% | 6.71% | 8.90% | 10.39% | -0.41% | 5.42% |
| Coffs Harbour Airport | 22.37% | 27.60% | 29.31% | 26.34% | 27.11% | 24.10% | 21.25% | 22.37% |
| Curtin Aero | 17.42% | 12.30% | 12.19% | 13.58% | 21.83% | 20.00% | 22.42% | 18.33% |
| Geraldton Airport | 21.88% | 25.51% | 24.72% | 25.24% | 24.91% | 21.30% | 20.31% | 21.67% |
| Hobart (Ellerslie Road) | 22.17% | 20.20% | 12.63% | 17.57% | 20.33% | 12.64% | 11.98% | 11.43% |
| Mount Isa | 9.52% | 10.41% | 13.59% | 17.48% | 8.35% | 7.77% | 12.37% | 18.17% |
| Tindal RAAF | -0.99% | -1.40% | 5.52% | 13.70% | 19.09% | 18.51% | 8.23% | 9.39% |
| Townsville | 40.44% | 40.21% | 39.99% | 40.68% | 34.52% | 29.96% | 33.72% | 38.71% |
| Woomera Aerodrome | -14.59% | -13.44% | 0.64% | 6.44% | 6.66% | 11.21% | 5.13% | 7.93% |

*Table A.2:* *Percentage of reduction of daily maximum RMSE by lead day, dew point temperature forecast.*

| Site/ Lead Time (Days) | 0 | 1 | 2 | 3 | 4 | 5 | 6 | 7 |
|---|---|---|---|---|---|---|---|---|
| Alice Springs Airport | 22.34% | 30.57% | 24.47% | 26.64% | 30.07% | 29.10% | 22.10% | 26.08% |
| Archerfield Airport | -22.87% | -24.99% | -21.63% | -13.30% | -16.71% | -6.05% | 3.43% | 4.45% |
| Avalon Airport | 10.76% | 17.61% | 19.01% | 16.96% | 21.53% | 19.75% | 17.86% | 21.69% |
| Coffs Harbour Airport | 11.97% | 11.58% | 10.07% | 14.37% | 12.35% | 14.26% | 18.65% | 17.65% |
| Curtin Aero | -4.10% | -0.74% | 5.55% | 11.45% | 12.07% | 11.54% | 14.79% | 13.29% |
| Geraldton Airport | 56.52% | 58.10% | 62.29% | 60.62% | 60.50% | 58.97% | 56.76% | 55.74% |
| Hobart (Ellerslie Road) | 8.77% | -5.70% | -2.54% | 0.98% | 2.58% | 1.53% | -8.63% | -6.81% |
| Mount Isa | -17.98% | -12.35% | -9.86% | -0.34% | 6.55% | 7.97% | 16.46% | 16.59% |
| Tindal RAAF | -8.02% | -14.47% | -9.09% | -6.19% | -2.40% | -4.88% | 5.70% | 4.93% |
| Townsville | 41.35% | 41.40% | 41.61% | 41.98% | 43.87% | 42.99% | 41.05% | 38.62% |
| Woomera Aerodrome | 2.22% | -0.21% | 1.63% | -0.62% | 2.67% | 9.03% | 5.14% | 7.87% |

*Table A.3:* *Percentage of reduction of daily minimum RMSE by lead day, temperature forecast.*

| Site/ Lead Time (Days) | 0 | 1 | 2 | 3 | 4 | 5 | 6 | 7 |
|---|---|---|---|---|---|---|---|---|
| Alice Springs Airport | 27.07% | 24.88% | 26.07% | 18.81% | 14.28% | 16.20% | 31.08% | 28.07% |
| Archerfield Airport | 28.32% | 22.11% | 23.56% | 22.77% | 24.67% | 23.85% | 24.89% | 21.16% |
| Avalon Airport | 16.67% | 12.56% | 13.00% | 17.94% | 14.91% | 17.68% | 19.25% | 13.74% |
| Coffs Harbour Airport | 33.92% | 32.81% | 36.97% | 32.35% | 34.03% | 33.46% | 34.86% | 37.92% |
| Curtin Aero | 9.45% | 10.37% | 14.83% | 11.45% | 12.66% | 14.54% | 19.71% | 23.85% |
| Geraldton Airport | 33.48% | 34.13% | 34.10% | 30.78% | 25.27% | 22.23% | 27.16% | 25.60% |
| Hobart (Ellerslie Road) | 4.28% | 5.19% | 3.68% | 8.67% | 7.37% | 10.29% | 6.01% | 1.66% |
| Mount Isa | 26.31% | 27.49% | 29.11% | 22.43% | 23.41% | 19.70% | 27.23% | 23.23% |
| Tindal RAAF | -0.58% | 2.67% | 8.03% | 11.79% | 5.07% | 5.06% | 24.70% | 21.63% |
| Townsville | 1.43% | 4.40% | 3.12% | 6.61% | 2.38% | 1.86% | 1.99% | 11.35% |
| Woomera Aerodrome | -10.90% | -6.77% | -4.91% | -10.26% | -11.21% | 2.01% | 3.22% | 0.93% |

*Table A.4:* Percentage of reduction of daily minimum RMSE by lead day, dew point temperature forecast.

| Site/ Lead Time (Days) | 0 | 1 | 2 | 3 | 4 | 5 | 6 | 7 |
|---|---|---|---|---|---|---|---|---|
| Alice Springs Airport | -1.01% | -1.94% | 5.53% | 15.99% | 14.57% | 17.94% | 12.72% | 9.59% |
| Archerfield Airport | 47.64% | 42.17% | 39.90% | 41.47% | 38.15% | 34.59% | 35.87% | 35.91% |
| Avalon Airport | 3.84% | 9.68% | 9.57% | 12.27% | 14.82% | 13.42% | 5.51% | 6.76% |
| Coffs Harbour Airport | 13.62% | 12.14% | 18.29% | 19.17% | 15.54% | 19.32% | 17.73% | 19.63% |
| Curtin Aero | -3.98% | 1.22% | 4.51% | 7.58% | 8.47% | 8.37% | 6.47% | -0.52% |
| Geraldton Airport | 21.68% | 8.46% | 9.49% | 4.84% | 3.64% | 9.64% | 6.36% | 12.70% |
| Hobart (Ellerslie Road) | 20.57% | 21.22% | 21.69% | 19.79% | 22.14% | 23.02% | 16.99% | 16.53% |
| Mount Isa | 22.57% | 27.18% | 34.10% | 33.14% | 32.53% | 29.46% | 28.11% | 31.96% |
| Tindal RAAF | 24.14% | 21.34% | 20.64% | 30.12% | 20.13% | 19.69% | 16.34% | 14.90% |
| Townsville | -6.72% | -1.34% | -1.47% | 0.54% | 5.46% | 14.01% | 9.25% | -2.09% |
| Woomera Aerodrome | 34.81% | 30.51% | 33.03% | 34.27% | 33.56% | 30.85% | 30.09% | 28.04% |

*Table A.5:* Reduction of critical error rates, hourly temperature forecast.

| Site/ Lead Time (Days) | 0 | 1 | 2 | 3 | 4 | 5 | 6 | 7 |
|---|---|---|---|---|---|---|---|---|
| Alice Springs Airport | 7.39% | 8.01% | 10.66% | 10.40% | 10.06% | 7.85% | 7.84% | 8.53% |
| Archerfield Airport | 3.91% | 4.50% | 6.15% | 6.01% | 6.90% | 7.87% | 8.94% | 8.42% |
| Avalon Airport | 1.32% | 1.25% | 1.46% | 1.27% | 2.73% | 2.06% | 2.25% | 2.11% |
| Coffs Harbour Airport | 5.37% | 6.18% | 6.86% | 6.60% | 6.63% | 5.64% | 5.03% | 5.16% |
| Curtin Aero | 0.82% | 0.89% | 1.07% | 2.57% | 2.59% | 3.32% | 4.31% | 6.08% |
| Geraldton Airport | 16.50% | 16.42% | 15.31% | 15.83% | 14.62% | 14.42% | 12.49% | 12.87% |
| Hobart (Ellerslie Road) | 2.78% | 3.16% | 3.69% | 4.09% | 2.93% | 3.39% | 2.18% | 3.72% |
| Mount Isa | 2.73% | 1.46% | 3.21% | 3.76% | 3.88% | 3.22% | 3.42% | 5.10% |
| Tindal RAAF | 3.81% | 4.62% | 4.68% | 5.92% | 5.93% | 4.97% | 6.49% | 6.96% |
| Townsville | 2.58% | 3.67% | 4.08% | 4.59% | 4.39% | 4.22% | 5.02% | 5.68% |
| Woomera Aerodrome | 1.41% | 3.07% | 4.09% | 5.23% | 5.92% | 5.96% | 4.40% | 7.13% |

*Table A.6:* Reduction of critical error rates, hourly dew point temperature forecast.

| Site/ Lead Time (Days) | 0 | 1 | 2 | 3 | 4 | 5 | 6 | 7 |
|---|---|---|---|---|---|---|---|---|
| Alice Springs Airport | 4.46% | 6.27% | 6.77% | 8.83% | 8.02% | 8.40% | 7.71% | 8.09% |
| Archerfield Airport | 8.29% | 8.03% | 7.04% | 8.29% | 8.10% | 8.75% | 8.52% | 6.70% |
| Avalon Airport | 1.20% | 1.55% | 0.72% | 1.25% | 2.30% | 2.29% | 2.03% | 2.54% |
| Coffs Harbour Airport | 0.95% | 0.54% | 0.90% | 0.81% | 0.85% | 2.07% | 4.42% | 5.02% |
| Curtin Aero | 0.42% | 1.16% | 1.47% | 0.98% | -0.06% | 1.51% | 1.14% | 2.23% |
| Geraldton Airport | 20.26% | 21.60% | 22.20% | 23.98% | 25.51% | 23.37% | 24.28% | 23.24% |
| Hobart (Ellerslie Road) | 1.44% | 2.25% | 2.73% | 2.44% | 4.34% | 4.52% | 4.73% | 4.70% |
| Mount Isa | 3.44% | 3.88% | 4.88% | 4.76% | 4.48% | 5.46% | 8.77% | 11.27% |
| Tindal RAAF | 0.64% | 2.10% | 3.08% | 2.93% | 2.95% | 2.73% | 1.06% | 0.74% |
| Townsville | 1.84% | 3.09% | 4.22% | 6.39% | 8.75% | 9.48% | 9.38% | 9.66% |
| Woomera Aerodrome | 9.02% | 10.50% | 9.72% | 8.48% | 5.76% | 4.53% | 4.60% | 3.66% |